\documentstyle[eqsecnum,prd,aps,a4]{revtex}
\input epsf

\begin{document}
\title{Vorton Formation}
\author{C. J. A. P. Martins\thanks{Also at C. A. U. P.,
Rua do Campo Alegre 823, 4150 Porto, Portugal.
Electronic address: C.J.A.P.Martins\,@\,damtp.cam.ac.uk}
and
E. P. S. Shellard\thanks{Electronic address:
E.P.S.Shellard\,@\,damtp.cam.ac.uk}}
\address{Department of Applied Mathematics and
Theoretical Physics\\
University of Cambridge\\
Silver Street, Cambridge CB3 9EW, U.K.}
\maketitle

\begin{abstract}
In this paper we present the first analytic model
for vorton formation. We start by deriving the microscopic string
equations of motion in Witten's superconducting model, and show that
in the relevant chiral limit these coincide with the ones obtained from
the supersonic elastic models of Carter and Peter. We then numerically study
a number of solutions of these equations of motion and thereby suggest
criteria for deciding whether a given superconducting loop configuration
can form a vorton. Finally, using a recently developed model for the
evolution of currents in superconducting strings we conjecture, by comparison
with these criteria, that string networks formed at
the GUT phase transition should produce no vortons. On the other hand, a
network formed at the electroweak scale can produce vortons accounting for
up to $6\%$ of the critical density. Some consequences of
our results are discussed.
\end{abstract}
\pacs{98.80.Cq, 11.27.+d}

\section{Introduction}
\label{v-in}
As first pointed out by Witten \cite{witten}, cosmic strings can in
some circumstances (typically when the electromagnetic gauge invariance is
broken inside the string) behave as
`superconducting wires' carrying large currents and charges---up to the
order of the string mass scale in appropriate units.
The charge carriers can be either bosons or fermions (see \cite{vs} for a
review). The former type occurs when it becomes energetically
favourable for a charged Higgs field to have a
non-zero vacuum expectation value in the string core; the latter happens
when fermions couple to the string fields creating fermion zero modes.

It is well known that arbitrarily large currents are not allowed---there is a
critical value beyond which the current saturates. In other words, for large
enough winding number per unit length, the superconducting condensate is
quenched down, suppressing the current flow. Also, the current can decay by
magnetic flux-line tunnelling; this can be used to impose constraints on
allowed particle physics models.

If superconducting strings carry currents, they must also carry charges of
similar magnitude. This includes not only charges trapped at formation by
the Kibble mechanism but also the ones due to string inter-commuting
between regions of the string network with different currents. Just like
with currents, charge densities cannot have arbitrarily large magnitude---there
is a limit beyond which there will no longer be an energy barrier
preventing the charge carriers from leaving the string.

A rather important point is that the presence of charges on the string tends
to counteract  the current quenching effect discussed above. In fact, numerical
simulations of contracting string loops at fixed charge and winding number have
shown \cite{davsh} that a `chiral' state with equal charge and current
densities is approached as the loop contracts. In this limiting chiral
case, quenching is in fact eliminated completely.
This has several important consequences. Strings that have trapped charges as
a consequence of a phase transition can become superconducting even if the
formation of a condensate was otherwise energetically unfavoured. More
importantly, a string with both a charge and a current density will
have a non-zero angular momentum. 

In the cosmological context, these strings would of
course interact with the cosmic plasma, originating a number of
interesting consequences. The most remarkable of these, however,
has to do with the evolution of string loops.
If a superconducting string loop has an angular momentum, it is
semi-classically conserved, and it tries to resist the loop's
tension. This will at least increase the loop's lifetime. If the
current is too large, charge carriers will leave the string accompanied by
 a burst of
electromagnetic radiation, but otherwise it is possible that dynamically
stable loops form. These are called vortons \cite{vor}---they are stationary
rings that do not radiate classically, and at large distances they
look like point particles with quantised
charge and angular momentum. Their cosmological significance
comes from the fact that
they provide very strong constraints on allowed particle physics models,
since they behave like non-relativistic particles. According to current
belief \cite{vor,vor2}, if they are formed at
high enough energy scales they are as dangerous as magnetic monopoles,
producing an over-density of matter in disagreement with observations.
On the other hand, low-mass vortons could be a very interesting dark
matter candidate. Understanding the mechanisms behind formation and
evolution is therefore an essential cosmological task.

The overwhelming majority of the work done on cosmic strings so far was
concerned with the structureless Goto-Nambu strings (but see \cite{carternew}
and references therein for some exceptions). In the case of work on vortons,
this means that somewhat {\em ad-hoc} estimates had to be made for
some properties of the cosmic string network---notably for
microscopic quantities such as current and charge densities. This is despite
the fact
it has been recognised a long time ago that, even though they might be
computationally very useful \cite{ms,ms2,ms2a}, Goto-Nambu
models cannot realistically be expected to account for a number of
cosmologically relevant phenomena, due to the very limited number of degrees of
freedom available. Two such phenomena are the build-up of small-scale
structure and charge and current densities.

In this paper we fill this important gap by discussing the problem of
vorton formation in the context of the superconducting string models of
Witten \cite{witten} and of Carter and Peter \cite{wcp}
(sections \ref{v-wt} and \ref{v-cp}). Strangely enough, the issue of the
conditions for vorton formation has been so far neglected with respect to
those of their stability and cosmological consequences. We will start
by introducing these models and determining the microscopic string equations of
motion in each case. It will be shown that in the relevant chiral limit these
equations coincide---this also provides the first conclusive evidence of
the validity of the supersonic elastic models of Carter and Peter \cite{wcp}.

We then proceed to study the evolution of a number of loop solutions of these
equations numerically (sections \ref{v-fl} and \ref{v-ex}),
and from the results of this analysis parameters
will be introduced which characterise the loop's ability to evolve into a
vorton state (section \ref{v-vt}). Finally, we discuss a very simple
phenomenological model for the evolution of the superconducting
currents on the long cosmic string network \cite{mss}, based on the dynamics
of a `superconducting correlation length' (sections \ref{v-cr}--\ref{v-ff}).
Using this model we can therefore estimate the currents carried by string loops
formed at all relevant times, and thus (in principle) decide if
these can become
vortons (section \ref{v-rs}) and calculate the corresponding density
(section \ref{v-den}).

Based on our results, we don't expect any GUT vortons to form at all.
This is essentially because the friction-dominated epoch is very short for
GUT-scale strings \cite{ms}, so their currents and charges are never large
enough to prevent them from becoming relativistic---and therefore liable to
losses. Even if they did form, they wouldn't be in conflict with the
standard cosmological scenario if they decayed soon after the end of the
friction-domination epoch.

Hence we conclude that, in contrast with previously existing
estimates \cite{vor,vor2}, one cannot at the moment rule out GUT
superconducting string models. We should point out at the outset that there are
essentially three improvements in
the present work which justify the different end result for GUT-scale strings.
Firstly, by analysing simple (but physically relevant) loop solutions of the
microscopic string equations of motion for the Witten model, we can get a much
improved idea of how superconducting loops evolve and of how (and under which
conditions) they reach a vorton state. Secondly, by using a simple
model for the evolution of the currents on the long strings \cite{mss}
we can accurately determine the typical currents on each string loop at
the epoch of its formation. Finally, the use of the analytic formalism
previously introduced by the present authors \cite{ms,ms2a} allows us to
use a quantitative description throughout the paper, and in particular to
determine the loop sizes at formation.

As will become clear
below, when taken together these allow a detailed analysis of the process
of vorton formation to be carried out, either in the Witten
model (as is done in this paper) or any other that one considers relevant.
In contrast, note that Davis \& Shellard \cite{vor} restrict themselves to
the particular case of the initial Brownian Vachaspati-Vilenkin loops
with Kibble currents,
and do not consider the subsequent evolution of the network. On the other hand,
Brandenberger {\em et al.} \cite{vor2} make rather optimistic
order-of-magnitude estimates about the process of relaxation into a vorton
state. As it turns out, for high energy GUT scales, all these loops become
relativistic before reaching a vorton state.
Finally, neither of these treatments has the benefit of a quantitative
model for the evolution of the long-string network \cite{ms} which allows
one to accurately describe the process of loop production.

On the other hand, as we lower the string-forming energy
scale we expect more and more efficient vorton production, and the 'old'
scenario still holds. Therefore intermediate-scale superconducting strings
are still ruled out, since they would lead to a universe becoming
matter-dominated earlier than observationally allowed. Finally,
at low enough energy scales, vortons will be a dark matter candidate. For
example, for a string network formed around $T\sim10^2\,GeV$ (typical of
the electroweak phase transition) they can provide up to
$6\%$ of the critical density. A more detailed discussion of these issues
is left to a forthcoming publication \cite{inprep}.

Throughout this paper we will use fundamental units in which
$\hbar=c=k_B=Gm^2_{Pl}=1$.

\section{Witten's Microscopic Model}
\label{v-wt}
As first pointed out by Witten \cite{witten}, a low-energy
effective action for a superconducting string can
be derived in a way that is fairly similar to what is done in
the Goto-Nambu case
(see for example \cite{vs}). One has to adopt the additional assumptions
that the current is much smaller than the
critical current and that the electromagnetic vector potential $A_\mu$ is
slowly varying on the scale of the condensate thickness.

The derivation then proceeds as in the neutral case, except for the use of the
well-known fact that in two dimensions a conserved current can be written as the
derivative of a scalar field. One obtains
\begin{eqnarray}
S &=&\int\sqrt{-\gamma}\left[-\mu_0+\frac{1}{2}\gamma^{ab}\phi_{,a}\phi_{,b}
-qA_\mu x^\mu_{,a}\frac{\epsilon^{ab}}{\sqrt{-\gamma}}\phi_{,b}\right]d^2\sigma \\
& &-\frac{1}{16\pi}\int d^4x\sqrt{-g}F_{\mu\nu}F^{\mu\nu}
\, ; \label{witact}
\end{eqnarray}
the four terms are respectively the usual Goto-Nambu term, the inertia of the
charge carriers, the current coupling to the electromagnetic potential and
the external electromagnetic field ($\epsilon^{ab}$ is the alternating
tensor); note that this applies to both
the bosonic and the fermionic case \cite{vs}.

Recalling the usual definitions
\begin{equation}
A_a=x^\mu_{,a}A_\mu\, ,\label{defaa}
\end{equation}
\begin{equation}
F_{ab}=F_{\mu\nu}x^\mu_{,a}x^\nu_{,b}=A_{b,a}-A_{a,b}\, ,\label{deffmn}
\end{equation}
and defining $\Upsilon_{ab}$ to be the stress-energy tensor
of the scalar field $\phi$
\begin{equation}
\Upsilon_{ab}=\phi_{,a}\phi_{,b}-\frac{1}{2}\gamma_{ab}\phi_{,c}\phi^{,c}
\, ,\label{phiset}
\end{equation}
and the conserved current $J^a$ as
\begin{equation}
J^a=q\frac{\epsilon^{ab}}{\sqrt{-\gamma}}\phi_{,b}\, ,\label{conscurr}
\end{equation}
we can obtain the following equations of motion by varying the action
(\ref{witact}) with respect to $A_\mu$, $\phi$ and $x^\mu$ respectively
\begin{equation}
F^{\mu\nu}_{;\nu}\equiv-4\pi q j^\mu=-4\pi\int d^2\sigma\epsilon^{ab}
x^\mu_{,a}\phi_{,b}\delta^4\left(x-x(\sigma^a\right)\, ,\label{witeqfmn}
\end{equation}
\begin{equation}
\partial_a\left(\sqrt{-\gamma}\gamma^{ab}\phi_{,b}\right)+\frac{1}{2}q
\sqrt{-\gamma}{\tilde\epsilon}^{ab}F_{ab}=0\, ,\label{witeqphi}
\end{equation}
and
\begin{eqnarray}
\partial_a\left[\sqrt{-\gamma}\left(\gamma^{ab}+\frac{1}{\mu_0}
\Upsilon^{ab}\right)x^\alpha_{,b}\right] &+&\sqrt{-\gamma}
\left(\gamma^{ab}+\frac{1}{\mu_0}\Upsilon^{ab}\right)\Gamma^\alpha_{\sigma\rho}
x^\sigma_{,a}x^\rho_{,b}= \\
&=&\sqrt{-\gamma}(F^\alpha+f^\alpha)
\, , \label{witeqxmn}
\end{eqnarray}
where $\Gamma^\alpha_{\mu\nu}$ is the usual metric connection,
\begin{equation}
\Gamma^\alpha_{\mu\nu}=\frac{1}{2}g^{\alpha\lambda}
\left(g_{\alpha\mu,\nu}+g_{\nu\alpha,\mu}-g_{\mu\nu,\alpha}\right)
\, , \label{connectmtwo}
\end{equation}
and $F^\alpha$ is the Lorentz force
\begin{equation}
F^\alpha=\frac{1}{2\mu_0}F^\alpha_\nu x^\nu_{,a}J^a\, .\label{lorentzfor}
\end{equation}
We have also included the friction force
\begin{equation}
f^\alpha=\frac{1}{\ell_{\rm f}}\left(u^\alpha-x^\alpha_{,a}x^{\sigma,a}u_\sigma
\right)\, ,\label{frictzfor}
\end{equation}
using the same procedure as described in\cite{v1,ms}.
As shown in \cite{mss}, plasma effects are subdominant, except possibly in
the presence of background magnetic fields---either of `primordial' origin
or generated (typically by a dynamo mechanism) once proto-galaxies have formed.
Hence one
expects Aharonov-Bohm scattering \cite{rsa} to be the dominant effect,
and consequently we have \cite{ms}
\begin{equation}
\ell_{\rm f}=\frac{\mu}{\beta T_b^3}\, ,
\label{frilen}
\end{equation}
where $T_b$ is the background temperature and $\beta$ is a numerical factor
related to the number of particle species interacting with the string.

The effect of self-inductance leads to the renormalisation of both the
electromagnetic coupling and the scalar field $\phi$. Now, it is
well known that the Maxwell-Faraday
tensor includes both the external field and the field produced by
the string itself, but it can be shown that if one follows this renormalisation
procedure one can identify it with its external component, which henceforth
we assume to vanish \cite{vs}.

As we already pointed out, when dealing with superconducting
string loops we are essentially interested in the chiral
limit of this model, that is
\begin{equation}
\gamma_{ab}J^a J^b=0\, ,\label{chiraldef}
\end{equation}
or equivalently
\begin{equation}
\phi'{}^2=\epsilon^2 {\dot\phi}^2\, .\label{chiralcon}
\end{equation}
In this limit, introducing the simplifying function $\Phi$ defined as
\begin{equation}
\Phi(\chi)=\frac{{\dot\phi}^2}{\mu_0 \gamma_{00}}\, ,\label{defphi}
\end{equation}
and choosing the standard gauge conditions
\begin{equation}
\sigma^0=\tau\, , \qquad {\bf {\dot x}}\cdot{\bf x}'=0 \, ,
\label{newgauge}
\end{equation}
(with dots and primes respectively denoting derivatives with respect to the
time-like and space-like coordinates on the worldsheet as usual)
the string equations of motion in an FRW background with the line element
\begin{equation}
ds^2=a^2\left(d\tau^2-{\bf{dx}}^2\right)\, \label{newlel}
\end{equation}
(which implies that $\gamma_{00}=a^2(1-{\bf {\dot x}}^2)$)
have the form
\begin{equation}
\left[\epsilon\left(1+\Phi
\right)\right]{\dot{}} +\frac{\epsilon}{\ell_d}{\dot {\bf x}}^2 =
\Phi'-2\frac{{\dot a}}{a}\epsilon\Phi \, , \label{witchirti}
\end{equation}
and
\begin{equation}
\epsilon\left(1+\Phi\right) {\ddot {\bf x }}+\frac{\epsilon}{\ell_d}
(1-{\dot {\bf x}}^2) {\dot {\bf x}}=\left[\left(1-\Phi\right)
\frac{{\bf x}'}{\epsilon}\right]'+
\left({\dot \Phi}+2\frac{{\dot a}}{a}\Phi\right){\bf x}'
+2 \Phi{\dot{\bf x}}' \, , \label{witchirsp}
\end{equation}
where for simplicity we have introduced the `damping length'
\begin{equation}
\frac{1}{\ell_d}=a\left(2H+\frac{1}{\ell_{\rm f}}\right)\, .\label{defdaml}
\end{equation}
Finally, the worldsheet charge and current densities are respectively given by
\begin{equation}
\rho_{w}=q\epsilon{\dot \phi}\, ,\label{dwscg}
\end{equation}
and
\begin{equation}
j_{w}=q\frac{\phi'}{\epsilon}\, .\label{dwscr}
\end{equation}

Note that the Witten action is `microscopic' in the sense of being built using
only the properties of the underlying particle physics model \cite{witten}.
In the next section we will analyse the equations of motion obtained
form the action for the elastic supersonic models of Carter and Peter \cite{wcp},
which is is this sense `macroscopic'.

\section{Supersonic Elastic Models}
\label{v-cp}
In order to account for phenomena such as the build-up of charge and
current densities on cosmic strings, one must introduce additional degrees
of freedom on the string worldsheet. One such class of models, originally
introduced by Carter and co-workers is usually referred to as elastic
models (see \cite{carternew} and references therein, on which the following
two subsections are based).

\subsection{Basics of elastic models}
In general, elastic string models can be described by a Lagrangian
density depending on the spacetime metric $g_{\mu\nu}$,
background fields such as a Maxwellian-type gauge potential $A_\mu$ or
a Kalb-Ramond gauge field $B_{\mu\nu}$ (but not their gradients) and any
relevant internal fields (that will be discussed in detail
below). Note that the Goto-Nambu model has a constant Lagrangian density,
namely
\begin{equation}
{\cal L}_{GN}=-\mu_0  \, .
\label{gnambact}
\end{equation}
Upon infinitesimal
variations in the background fields, and provided that independent
internal fields are kept fixed (or alternatively that
their dynamic equations of motion
are satisfied), the action will change by
\begin{equation}
\delta S=-\frac{1}{2}\int \left(T^{\mu\nu}\delta g_{\mu\nu}+
W^{\mu\nu}\delta B_{\mu\nu}+2J^\mu A_\mu\right)
\sqrt{-\gamma}d^2\sigma  \, ,
\label{elactiovar}
\end{equation}
where
\begin{equation}
T^{\mu\nu}=2\frac{\delta{\cal L}}{\delta g_{\mu\nu}}+{\cal L}\eta^{\mu\nu} \,
\label{edeftmn}
\end{equation}
is the worldsheet the stress-energy tensor density,
\begin{equation}
J^\mu=\frac{\delta{\cal L}}{\delta A_\mu}\,
\label{edefemgc}
\end{equation}
is the worldsheet electromagnetic current density, and
\begin{equation}
W^{\mu\nu}=2\frac{\delta{\cal L}}{\delta B_{\mu\nu}}\,
\label{edefvort}
\end{equation}
is the worldsheet vorticity flux. The later
will not be considered further in this paper.

It useful to define two
orthogonal unit vectors tangent to the worldsheet, one of them
being time-like and the other space-like, such that
\begin{equation}
-U^\mu U_\mu =V^\mu V_\mu=1 \, , \qquad U^\mu V_\mu=0  \, .
\label{reluv}
\end{equation}
The eigenvalues of this ortho-normal frame are the energy density in
the locally preferred string rest frame, which will henceforth be denoted
by $U$, and the local
string tension, denoted $T$ (there should be no confusion with the
vector $U^\mu$ defined in (\ref{reluv}) and the stress-energy tensor
$T^{\mu\nu}$ defined in (\ref{edeftmn}), respectively), so that one can write
\begin{equation}
T^{\mu\nu}=UU^\mu U^\nu -T V^\mu V^\nu \, .
\label{elasttut}
\end{equation}
Note that $U$ and $T$ are simply constants for a Goto-Nambu string,
\begin{equation}
U=T=\mu_0  \, ,
\label{utgotonamb}
\end{equation}
but they are variable in general---hence the name `elastic strings'.
In particular, one should expect that the string tension in an elastic
model will be reduced with respect to the Goto-Nambu case due to the
mechanical effect of the current.

Since elastic string models necessarily possess conserved currents, it is
convenient to define a `stream function' $\psi$ on the worldsheet that
will be constant along the current's flow lines. The part of the
Lagrangian density
${\cal L}$ containing the internal fields is usually called the
`master function', and can be defined as a function of the magnitude of the
gradient of this stream function, $\Lambda=\Lambda(\chi)$,
such that
\begin{equation}
\chi=\gamma^{ab}\psi_{;a} \psi_{;b}  \, ,
\label{eldefchi}
\end{equation}
where the gauge covariant derivative is defined as
\begin{equation}
\psi_{;a}=\partial_a \psi-eA_\mu x^\mu_{,a}  \, .
\label{eldefcogder}
\end{equation}
Note that the definition of $\chi$ differs by a minus sign from that of
Carter \cite{carternew}; the reason for this will become clear below.
This `dynamic' term contains charge couplings,
whose relevance will be further discussed below. Nevertheless, whether or
not these or other background gauge fields are present, it is always the
form of the master function which determines the equation---or equations---of
state.

There is also a dual\cite{dual} potential ${\tilde \psi}$, whose gradient is
orthogonal to that of $\psi$, and the corresponding
dual master function ${\tilde \Lambda}={\tilde \Lambda}({\tilde \chi})$
such that
\begin{equation}
{\tilde \chi}=\gamma^{ab}{\tilde \psi}_{;a}{\tilde \psi}_{;b}  \, ,
\label{eldefchit}
\end{equation}
with the obvious definition for ${\tilde \psi}_{;a}$.
The duality between these descriptions means that the field equations
for the stream function $\psi$ obtained with the master function $\Lambda$
are the same as those for the dual potential ${\tilde \psi}$ obtained with
the dual master function ${\tilde \Lambda}$. However, there will in general
be two different equation of state relating the energy density $U$
and the tension $T$; these correspond
to what is known as the `magnetic' and `electric' regimes, respectively
corresponding to the cases
\begin{equation}
{\tilde \chi}_{mg}\, < \, 0 \, < \, \chi_{mg} \,  \label{elmag}
\end{equation}
and
\begin{equation}
\chi_{el}\, < \, 0 \, < \, {\tilde \chi}_{el} \, , \label{elele}
\end{equation}
that are respectively characterised by space-like and time-like currents.
In the degenerate null state limit, however, there will be a single equation,
\begin{equation}
U=T=\mu_0 \, . \label{nullcase}
\end{equation}
Note that the distinction between
a given model and its dual disappears in the absence of charge couplings;
such models are then called `self-dual' for obvious reasons.

In each case the equation of state provides the expressions
\begin{equation}
c^2_E=\frac{T}{U} \, , \label{cdspe}
\end{equation}
\begin{equation}
c^2_L=-\frac{dT}{dU}=\frac{\nu}{\mu}\frac{d\mu}{d\nu} \, , \label{cdspl}
\end{equation}
for the extrinsic (that is transverse, or `wiggle') and for the
sound-type (longitudinal or `woggle') perturbations of
the worldsheet. Both of these must obey
$c^2\ge0$ (a requirement for local stability) and $c^2\le1$ (a requirement
for local causality). These two speeds can be used to characterise the elastic
model in question; in particular there is a straightforward but
quite meaningful division of the models into supersonic (that is,
those obeying $c_E>c_L$), transonic ($c_E=c_L$; only in the null limit is
this common speed unity) and subsonic ($c_E<c_L$).

\subsection{Supersonic (superconducting) models}
Carter and Peter \cite{wcp} have recently proposed two supersonic
elastic models to describe the behaviour of current-carrying
cosmic strings. The Lagrangian density in the magnetic regime is
\begin{equation}
{\tilde \Lambda}_{mg}=-m^2+\frac{{\tilde \chi}}{2}
\left(1-\frac{{\tilde \chi}}{2k_0m^2_\sigma}\right)^{-1} \, , \label{wcpmag}
\end{equation}
$m_\sigma$ being the current carrier mass (which is at most of the order
of the relevant Higgs mass);
this is valid in the range
\begin{equation}
-\frac{1}{3}<\frac{{\tilde \chi}}{k_0m^2_\sigma}<
1-\frac{k_0m^2_\sigma}{2m^2+k_0m^2_\sigma} \, , \label{wcpmagran}
\end{equation}
and obeys the equation of state
\begin{equation}
\frac{U}{m^2}=1+\frac{k_0 m^2_\sigma}{4m^2}-\frac{\sqrt{k_0}}{\sqrt{2}}
\frac{m_\sigma}{m}\left(\frac{T}{m^2}-1+\frac{k_0 m^2_\sigma}{8m^2}\right)^{1/2}
 \, . \label{wcpmageqs}
\end{equation}
On the other hand, the electric regime is described by the Lagrangian density
\begin{equation}
{\tilde \Lambda}_{el}=-m^2-\frac{k_0m^2_\sigma}{2}
\ln{\left(1-\frac{{\tilde \chi}}{k_0m^2_\sigma}\right)} \, , \label{wcpel}
\end{equation}
which is valid in the range
\begin{equation}
-1<\frac{{\tilde \chi}}{k_0m^2_\sigma}<
1-e^{-2m^2/k_0m^2_\sigma} \, , \label{wcpelran}
\end{equation}
and the corresponding equation of state is
\begin{equation}
\frac{U}{m^2}=\frac{T}{m^2}+\frac{k_0m^2_\sigma}{m^2}
\left[\exp{\left(2\frac{1-T/m^2}{k_0m^2_\sigma/m^2}\right)}-1\right]
\, . \label{wcpeleqs}
\end{equation}
These models are supersonic for all space-like, and weak time-like
currents, with the exception that in the null limit ${\tilde \chi}=0$ one has
$c_L=c_E=1$.

\subsection{Equations of motion}
We now derive the microscopic equations of motion for elastic cosmic
string models. It is convenient to start by defining the quantity
\begin{equation}
\Theta^{ab}\equiv {\tilde \Lambda}\gamma^{ab}-2
\frac{\partial {\tilde \Lambda}}{\partial {\tilde \chi}}
{\tilde \psi}^{;a}{\tilde \psi}^{;b}\, ; \label{dfttt}
\end{equation}
then recalling the definition of ${\tilde \chi}$,
(\ref{eldefchit}), one can find the free string equations of motion in the
usual (variational) way, obtaining
\begin{equation}
(\sqrt{-\gamma}\Theta^{ab}x^\alpha_{,b})_{,a}+\sqrt{-\gamma}
\Theta^{ab}\Gamma^{\alpha}_{\mu\nu}x^{\mu}_{,a}x^{\nu}_{,b}=0 \, .
\label{xmneqnfree}
\end{equation}

Also in a similar way to what was done in section \ref{v-wt}, the effect
of the frictional forces is accounted for by introducing a term
$\sqrt{-\gamma}F^\alpha$ on the right-hand side of (\ref{xmneqnfree}).
For exactly the same reasons as those of section, we
will have
\begin{equation}
F^\alpha=-\beta T_b^3\left(u^\alpha-x^\alpha_{,a}x^{\beta,a}u_\beta\right)
\, . \label{deffricel}
\end{equation}
However, it should be remarked that the generalised definition of
the friction lengthscale for elastic models is
\begin{equation}
\ell_{\rm f}=-\frac{{\tilde \Lambda}}{\beta T_b^3}\,  \label{newfrl}
\end{equation}
(note that ${\tilde \Lambda}$ is negative).

Of course we now have a further equation for the scalar field
${\tilde \psi}$, namely
\begin{equation}
\partial_a\left(\sqrt{-\gamma}
\frac{\partial{\tilde \Lambda}}{\partial{\tilde \chi}}
\gamma^{ab}{\tilde \psi}_{;b}\right)=0\, . \label{neweqpsi}
\end{equation}

Furthermore, the spacetime energy-momentum tensor and
electromagnetic current will be given by
\begin{equation}
\sqrt{-g}T^{\mu\nu}=-\int
\sqrt{-\gamma}\Theta^{ab}x^\mu_{,a}x^\nu_{,b}\delta\left(x-x(\sigma,\tau)
\right)d^2\sigma\, \label{elspem}
\end{equation}
and
\begin{equation}
\sqrt{-g}J^\mu=-2e\int
\sqrt{-\gamma}\gamma^{ab}\frac{\partial{\tilde \Lambda}}{\partial{\tilde \chi}}
{\tilde \psi}_{,a}x^\mu_{;b}\delta\left(x-x(\sigma,\tau)
\right)d^2\sigma\, . \label{elspcr}
\end{equation}
The total string energy and charge in a spacetime where the line element
is (\ref{newlel}) are then defined as (we are again using the gauge
choice (\ref{newgauge}))
\begin{equation}
E=\int {\bf{d^3x}}\sqrt{-^{(3)}g}T^0_0=
a\int\left(-{\tilde \Lambda}+2
\frac{\partial{\tilde \Lambda}}{\partial{\tilde \chi}}
{\tilde \psi}_{;0}{\tilde \psi}^{;0}\right)\epsilon d\sigma\, \label{defen}
\end{equation}
and
\begin{equation}
Q=\int {\bf{d^3x}}\sqrt{-^{(3)}g}J_0=
-2ea\int\frac{\partial{\tilde \Lambda}}{\partial{\tilde \chi}}
{\tilde \psi}_{;0}\epsilon d\sigma\, .\label{defch}
\end{equation}
The corresponding worldsheet charge and current densities are defined via
\begin{equation}
j^\mu=-2e\frac{\partial{\tilde \Lambda}}{\partial{\tilde \chi}}\sqrt{-\gamma}
\gamma^{ab}
{\tilde \psi}_{,a}x^\mu_{,b}\equiv x^\mu_{,b}j^b\, ,\label{defedshcurr}
\end{equation}
and have the following values
\begin{equation}
\rho\equiv j^0=-2e\frac{\partial{\tilde \Lambda}}{\partial{\tilde \chi}}
\epsilon {\tilde \psi}_{;0}\, ,\label{wdshrho}
\end{equation}
\begin{equation}
j\equiv -j^1=-2e\frac{\partial{\tilde \Lambda}}{\partial{\tilde \chi}}
\frac{{\tilde \psi}_{;1}}{\epsilon}\, .\label{wdshjj}
\end{equation}

Again, for the reasons explained above, a particularly relevant situation will
be that of a chiral current, that is one in which
\begin{equation}
\gamma_{ab}j^aj^b=0\, .\label{chir}
\end{equation}
This is equivalent to
\begin{equation}
{\tilde \psi}'{}^2=\epsilon^2{\dot {\tilde \psi}}^2\, ,\label{chirpsi}
\end{equation}
and therefore it implies that
\begin{equation}
\frac{\partial\rho}{\partial\tau}=\frac{\partial j}{\partial\sigma}
\, \label{chirj}
\end{equation}
and that the total (spacetime) charge and current are also equal. Note that
in the chiral case one also has
\begin{equation}
{\tilde \chi}=0\, , \qquad
2\frac{\partial{\tilde \Lambda}}{\partial{\tilde \chi}}=1
\, ,\label{chirlagr}
\end{equation}
so this is not equivalent to
the Goto-Nambu case despite the fact that the equation of state is
\begin{equation}
U=T=\mu_0\, .\label{chireqn}
\end{equation}

Last but not least, one can always define the fundamental
worldsheet current density vector
\begin{equation}
i^a=-\sqrt{-\gamma}\frac{\partial{\tilde\Lambda}}{\partial{\tilde\psi}_{;a}}
=2\sqrt{-\gamma}\frac{\partial{\tilde\Lambda}}{\partial{\tilde\chi}}
\gamma^{ab}{\tilde\psi}_{;b}\, .\label{funcurra}
\end{equation}

Although we have included the charge coupling term is the master function and
its dual, it should be said that charge coupling
effects are subdominant, and thus for most purposes they can be neglected
(if nothing else,  at least to a first-order approximation).
This has been confirmed by Peter \cite{ppeter},
and is a consequence of the smallness of the coupling constants---for
example, the electromagnetic coupling constant is $e^2\sim1/137$.
In most of what follows we will therefore neglect the charge coupling.

If an electromagnetic coupling does exist, it will be simply given by
\begin{equation}
i^\mu=ej^\mu\, ,\label{relcoups}
\end{equation}
where
\begin{equation}
i^\mu=x^\mu_{,a}i^a\, ,\label{obvio}
\end{equation}
is the corresponding spacetime current. Note that (\ref{neweqpsi}) is then just
a Noether identity,
\begin{equation}
i^a_{;a}=0\, .\label{noethr}
\end{equation}

\subsection{The chiral limit}
We now consider the (common) chiral limit of the two supersonic elastic
models of Carter and Peter \cite{wcp}, defined by the Lagrangian densities
(\ref{wcpmag}) and (\ref{wcpel}), respectively for the magnetic and electric
regimes.
Also, as we did in section \ref{v-wt} for the Witten model, we will
interpret the charge coupling and the scalar field as being renormalised
and neglect the coupling to external electromagnetic fields.

Then, with our usual gauge choices and definitions of the damping and
friction length-scales, the microscopic string equations of motion
(\ref{xmneqnfree}) simplify to
\begin{equation}
\left[\epsilon\left(1+\Psi
\right)\right]{\dot{}} +\frac{\epsilon}{\ell_d}{\dot {\bf x}}^2 =
\Psi'-2\frac{{\dot a}}{a}\epsilon\Psi \, , \label{witchirtiww}
\end{equation}
and
\begin{equation}
\epsilon\left(1+\Psi\right) {\ddot {\bf x }}+\frac{\epsilon}{\ell_d}
(1-{\dot {\bf x}}^2) {\dot {\bf x}}=\left[\left(1-\Psi\right)
\frac{{\bf x}'}{\epsilon}\right]'+
\left({\dot \Psi}+2\frac{{\dot a}}{a}\Psi\right){\bf x}'
+2 \Psi{\dot{\bf x}}' \, , \label{witchirspww}
\end{equation}
where $\Psi$ is defined as
\begin{equation}
\Psi({\tilde \psi})=\frac{{\dot {\tilde \psi}}^2}{\mu_0 \gamma_{00}}
\, .\label{defpsi}
\end{equation}

That is, these are exactly the same equations of motion as those of
Witten's model (\ref{witchirti}--\ref{witchirsp}) if one identifies
the corresponding scalar fields,
\begin{equation}
\phi\equiv{\tilde \psi}\, .\label{ident}
\end{equation}

Then, the worldsheet charge and current densities also coincide,
\begin{equation}
\rho_{w}=q\epsilon{\dot{\tilde \psi}}\, ,\label{dwscgww}
\end{equation}
\begin{equation}
j_{w}=q\frac{{\tilde \psi}'}{\epsilon}\, .\label{dwscrww}
\end{equation}
Finally, the total energy of a piece of string is given by
\begin{equation}
E=\mu_0 a\int\left(1+\Psi\right)\epsilon d\sigma\, ,\label{chirtoten}
\end{equation}
which we can immediately interpret as being split in an obvious way
into a `string' component and a `current' component. This interpretation will
be relevant below.

Note that if we had preserved Carter's original sign conventions we
would have found a difference of a factor of $i$ between the two
fields. But the important point is that the equality between the
two theories in the chiral limit is not entirely trivial since, as we already
pointed out, the motivations behind the build up of each of them are quite
different. We have thus provided the first substantive evidence of the validity
of the supersonic elastic models of Carter and Peter \cite{wcp}.

\section{Chiral loops in flat spacetime}
\label{v-fl}
We will now study the evolution of current-carrying cosmic string loops,
starting by considering the simplest case of circular
loops in flat spacetime. We therefore choose the {\em ansatz}
\begin{equation}
{\bf x}(\tau)=r(\tau)(\sin\theta,\cos\theta,0)\, ;\label{ansatzag}
\end{equation}
we also need an ansatz for the scalar field ${\tilde \psi}$ (or $\phi$),
which we will take to be
\begin{equation}
{\tilde \psi}=\sqrt{\mu_0}({\cal F}(\tau)+n\sigma)t_c\, ,\label{ansatznew}
\end{equation}
where the winding number per unit $\sigma$, $n$, is a constant (due to
the symmetry of our loop solution) and $t_c$ is a characteristic
timescale---say the epoch of network formation.
The chirality condition implies that
\begin{equation}
\epsilon{\dot{\cal F}}=n\, .\label{chcdimp}
\end{equation}
Then the string equations of motion reduce to
\begin{equation}
\epsilon\left(1+\frac{n^2t_c^2}{r^2}\right)=1\, ,\label{chfl1}
\end{equation}
\begin{equation}
{\ddot r}+\left(1-\frac{n^4t_c^4}{r^4}\right)r=0\, ,\label{chfl2}
\end{equation}
together with the constraint
\begin{equation}
|n|\le\frac{1}{2}\, .\label{flatnconstr}
\end{equation}
Note that opposite signs of $n$ correspond to left and right moving currents;
naturally it always appears as $n^2$ in any relevant equation,
and we will therefore be taking $n$ to be positive.

\begin{figure}
\vskip-0.5in
\vbox{\centerline{
\hskip-3em\epsfxsize=.4\hsize\epsfbox{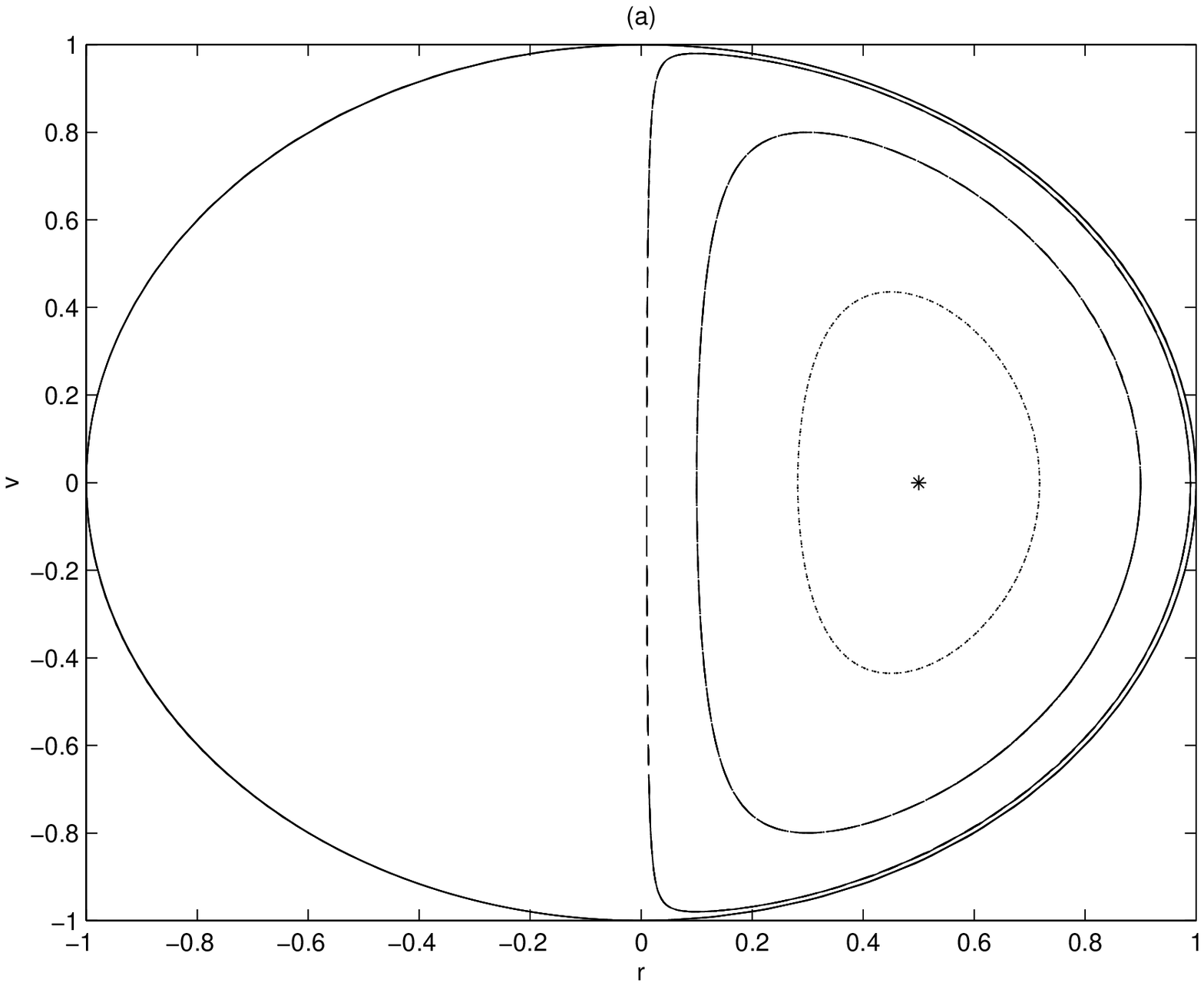}
\hskip5em\epsfxsize=.4\hsize\epsfbox{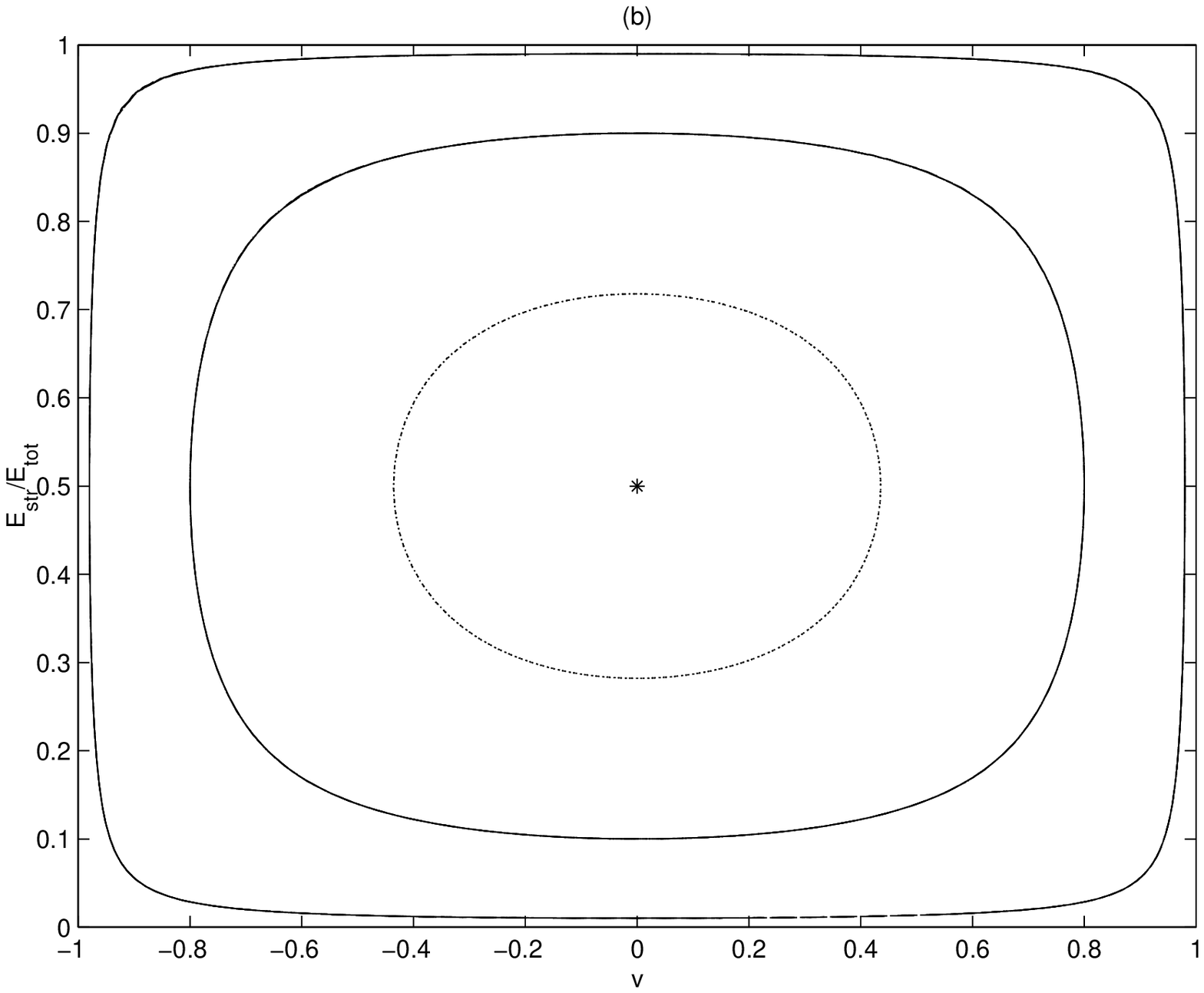}}
\vskip0.2in}
\caption{The flat spacetime evolution of chiral circular string
loops characterised by a conserved quantity $n$ (defined in
\protect\ref{ansatznew}) having the value $0$ (solid lines), $0.1$ (dashed),
$0.3$ (dot-dashed), 0.45 (dotted) and 0.5 (star).
Note that the first corresponds to a simple Goto-Nambu loop, while the last is
a static solution. Plots respectively show  ${\protect\dot r}$ as function
of $r$ (defined in \protect\ref{ansatzag}) (a) and the fraction of
the energy in the string as a function of ${\protect\dot r}$ (b).}
\label{chirflat}
\end{figure}

In figure \ref{chirflat} we plotted some relevant evolutionary properties of
chiral superconducting loops with different $n$'s in flat spacetime. Note that
these loops never collapse to zero size, and that their microscopic
velocity is always less than unity (unlike in the Goto-Nambu case).
Furthermore, there is a static solution with
\begin{equation}
n=\frac{1}{2}=\frac{r}{t_c}\, \qquad {\dot r}=0 \, ;\label{fstatic}
\end{equation}
in this case the energy is equally divided between the string and the current.

It should also be noted that energy
is transferred back and forth between the string and the current as the loop
oscillates. We can easily determine the following quantities (the averages
are over one oscillation period)
\begin{equation}
\langle \frac{r^2}{t_c^2}\rangle=\frac{1}{2}-n^2\, ,\label{flavr2}
\end{equation}
\begin{equation}
\langle\frac{t_c^2}{r^2}\rangle=\frac{1}{n^2}\, ,\label{flav1r}
\end{equation}
\begin{equation}
\langle {\dot r}^2\rangle=\frac{1}{2}(1-4n^2)\, ,\label{flavrd}
\end{equation}
while the energy the string obeys
\begin{equation}
\langle \frac{E_{string}}{E_{total}}\rangle=1-n\, ,\label{fbigr1}
\end{equation}
\begin{equation}
\langle \frac{E_{string}^2}{E_{total}^2}\rangle
=1-\frac{3}{2}n\, ;\label{fbigr2}
\end{equation}
note that the energy of these loops is $E_{total}/t_c=2\pi\mu$.

Finally, two other points that will have further relevance below. Firstly, a
loops with a given conserved number $n$ will reach a maximum
microscopic velocity (and corresponding Lorentz factor) given by
\begin{equation}
{\dot r}^2_{max}=1-4n^2\, ,\qquad \gamma_{max}=\frac{1}{2n}.\label{flmaxv}
\end{equation}
Secondly, for fixed $n$ and initial velocity, there will be two possible
choices of $r_i$ that can be made---the difference is that in one of them
most of the energy will be in the string, while in the other it will be in
the current. We will call these two cases the `string branch' and the `current
branch'. In flat spacetime, the two choices give physically the same solution
(they simply correspond to different initial phases of the oscillation),
but this will not be true in general.

\section{Chiral loops in the expanding universe}
\label{v-ex}
The case of circular loops in expanding universes is analogous, and we keep the
{\em ans\"atze} for ${\bf x}$ and ${\tilde \psi}$, 
\begin{equation}
{\tilde \psi}=\sqrt{\mu_0}({\cal F}(\tau)+n\sigma)t_c\, .\label{ansatznew2}
\end{equation}
The winding number per unit $\sigma$ and the function ${\cal F}$ are also
constrained as before. In terms of these quantities the total energy
of the loop can be written as
\begin{equation}
E_{total}=\mu_0\ell_{total}=\mu_0a\int\left(1+\frac{n^2t^2_c}{a^2 r^2}\right)
\epsilon d\sigma\equiv\mu_0\ell_{string}+E_{current}\, ,\label{chirtoexp}
\end{equation}
and the loops evolve according to
\begin{equation}
\left(1+\frac{n^2t^2_c}{a^2r^2}\right){\ddot r}+(1-{\dot r}^2)
\left[\frac{{\dot r}}{\ell_d}+\left(1-\frac{n^2t^2_c}{a^2r^2}\right)
\frac{1}{r}\right]=0\, .\label{chflexp}
\end{equation}

It is convenient to define a macroscopic dimensionless parameter which,
as we will show later, turns out to measure the loop's stability against
collapse. We will define it by
\begin{equation}
{\overline n}=\frac{4\pi nt_c}{\ell_{total}}\, ;\label{ansatznbar}
\end{equation}
note that unlike $n$ which is a constant for each loop,
${\overline n}$ is a variable parameter obeying
\begin{equation}
0\le {\overline n}\le 1\, ;\label{limnbar}
\end{equation}
also ${\overline n}=0$ corresponds to the Goto-Nambu case, while the
${\overline n}=1$ limit is the analogous of the flat spacetime
static solution, here characterised by
\begin{equation}
E_{string}=E_{current}\, ,\qquad {\dot r}=0 \, ;\label{reslnbar}
\end{equation}
in the approach to this limit one can easily establish that the loop's
velocity (in the radiation epoch) and length in string evolve according to
\begin{equation}
v\frac{t}{t_c}=\frac{n}{2}\, ,\qquad \frac{\ell_{string}}{2\pi t_c}=
\frac{n}{\sqrt{1-{\dot r}^2}} \, ;\label{vellnbar}
\end{equation}
these will be numerically confirmed below.

\begin{figure}
\vbox{\centerline{
\epsfxsize=0.6\hsize\epsfbox{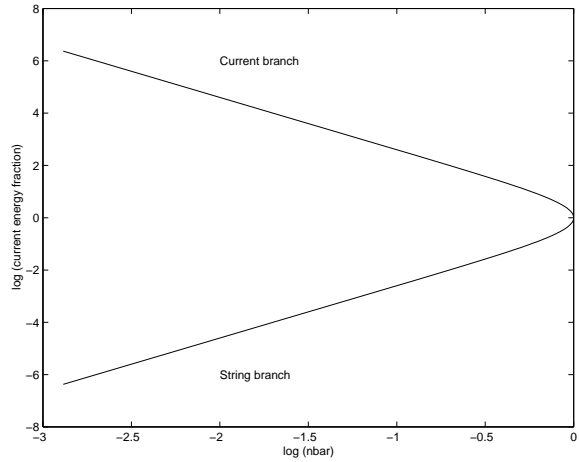}}
\vskip.4in}
\caption{The logarithm of the ratio of the energies
in the current and in the string for chiral circular loops with zero
velocity, as a function of the logarithm
of parameter ${\protect\overline n}$. Note that there are two different
branches, hereafter called the `current branch' (top) and the `string
branch' (bottom).}
\label{figbra}
\end{figure}

An important difference with respect to the flat spacetime case is that
now the string branch and the current branch (see figure \ref{figbra} for a
relevant particular case) represent two physically different
solutions---something to be expected since damping forces (that is,
friction and expansion) act differently on the string and
current energies. Since we will be mostly interested in chiral superconducting
string loops formed in the friction-dominated regime (as no vortons will
form in the `free' regime), we can safely assume
that these loops are formed with zero velocity. Now, there is a very simple
relation between $\ell_{total}$, $\ell_{string}$ and ${\overline n}$, namely
\begin{equation}
\frac{\ell_{total}}{\ell_{string}}=\frac{2}{{\overline n}^2}\left(1\pm
\sqrt{1-{\overline n}^2}\right) \, .\label{llsnbar}
\end{equation}
The negative sign corresponds to the string branch, where as ${\overline n}$
goes from zero to unity we go from the Goto-Nambu case to the static case
where the energy is split equally between the string and the current; the
positive sign corresponds to the current branch, where the ratio of the
energies in the string and in the current decreases until it vanishes when
${\overline n}$ reaches zero again. Note that (\ref{llsnbar}) can be inverted
to give
\begin{equation}
{\overline n}=2\frac{\ell_{string}}{\ell_{total}}\left(
\frac{\ell_{total}}{\ell_{string}}-1\right)^{1/2}\, .\label{llsnbar2}
\end{equation}

In practice, it is not easily conceivable that in cosmological contexts loops
can be formed with more energy in the current than in the string itself.
Therefore, although for the sake of completeness
we will be discussing the current branch in the
remainder of this section, we will neglect it afterwards.

Thus from (\ref{chflexp}) one obtains the evolution equation for
$\ell(\ell_i,t_i,{\overline n}_i,t)$, $v(\ell_i,t_i,{\overline n}_i,t)$
and other relevant quantities. As we will see below, a crucial quantity will
be the the maximum velocity reached by each loop configuration during it
evolution, $v_{max}(\ell_i,t_i,{\overline n}_i)$. If the loop does become
a vorton, than its length will asymptotically be given by
$\ell_v(\ell_i,{\overline n}_i)$.

In figures \ref{chirrad}--\ref{chirsmall} we plot the cosmological evolution of
some relevant GUT-scale chiral circular loops. We should mention that
in order to save space, only one out of every forty points resulting from the
numerical integrations is plotted, and this is the reason why some plots show
irregularities.

Figure \ref{chirrad} shows some relevant properties of
the evolution of chiral circular GUT-scale loops
formed at $t=t_c$; all have an initial total energy $E_{total}/2\pi\mu t_c=10$,
but the distribution of the energy between the string and the current varies.

Obviously, loops with higher currents will have smaller physical radii, and
hence they will be less stretched by expansion and enter the horizon earlier,
at which point they start oscillating---as can be confirmed
in \ref{chirrad}(a-b).
Regarding the velocities, note the significant differences between loops
in the `string branch' (which still reach fairly high microscopic
velocities, but never $v=1$) and in the `current branch' (which quickly
become non-relativistic). Therefore the latter ones should definitely become
vortons, and so it is perhaps fortunate that, as we pointed out above, we
do not expect loops with such high currents to be produced in the early
universe (at least, for GUT-scale networks). Note that in one of the
cases shown
the initial current is so high that
the loop `overshoots' and acquires a fairly large velocity, but friction
quickly slows it down again.

On the other hand, in the string branch the velocity is reduced with respect to
the Goto-Nambu case, and a more detailed investigation will be needed to set
up some criterion defining which velocities will allow vorton
formation---recall
that relativistic velocities will imply charge losses and it will therefore
be unrealistic to make any definite claims or predictions about such cases.

The evolution of the fraction of the loop's energy in the current is
particularly illuminating (see \ref{chirrad}(c)). This will obviously decrease
while the loop is being stretched, and it will start oscillating when the loop
falls in side the horizon. The oscillations are around the state with
equipartition of the energy between the string and the current, which as
we saw corresponds to a static solution in flat spacetime. Note that the
effect of the friction force is to reduce the amplitude of these oscillations,
so one can see that friction is in fact crucial for vorton formation.
Naturally, loops with smaller velocities will undergo oscillations with smaller
amplitudes, so again we confirm that these are the strongest vorton candidates.
Finally, we have plotted the parameter ${\overline n}$
(which was defined in \ref{ansatznbar}) in \ref{chirrad}(d), and as
one can easily see by comparison with the other three plots this is indeed a
good indicator of whether or not a given loop can become a vorton---in fact,
the `phenomenological' criterion that we mentioned above will be basically
expressed in terms of the value of ${\overline n}$ once the loop is
`free'---that is, much smaller than the damping length defined
in (\ref{defdaml}).

On the other hand, radiative backreaction also tends to damp these energy
oscillations, and consequently increase ${\overline n}$. Note that this has
been shown to have the approximate form ${\dot E}=\Gamma_{em}J^2$, and since
$\Gamma{_em}\sim100$ the timescale for this process is expected to be
relatively short.

Note that when loops become smaller than the damping lengthscale
$\ell_{\rm d}$ and reach the `free' regime the following averages over a period
hold (note that ${\overline n}$ becomes a constant in this limit---hence
its usefulness)
\begin{equation}
\langle {\dot r}^2\rangle=\frac{1}{2}(1-{\overline n}^2)\, ,\label{flavrdn}
\end{equation}
\begin{equation}
\langle \frac{E_{string}}{E_{total}}\rangle=1-\frac{1}{2}{\overline n}
\, ,\label{fbigr1n}
\end{equation}
\begin{equation}
\langle \frac{E_{string}^2}{E_{total}^2}\rangle=1-\frac{3}{4}{\overline n}
\, ;\label{fbigr2n}
\end{equation}
the variance of the fraction of the energy in string is therefore
\begin{equation}
\Delta \frac{E_{string}}{E_{total}}=\frac{1}{4}{\overline n}(1-{\overline n})
\, .\label{fbigr2var}
\end{equation}

In figure \ref{chirfri} we show chiral loops with the same initial conditions
as \ref{chirrad}, but starting to evolve at the epoch $t_\star$ when
when friction becomes negligible \cite{ms}.
The differences are self-evident. Now, after a first period of
growth of the total radius due to expansion, there is no mechanism forcing
the loops to return this extra energy back to the medium when they fall
inside the horizon. Consequently
there is also no velocity damping (all loops will have microscopic velocities
larger than $0.5$) and the energy oscillations between the string and
the current always have a large amplitude---so that ${\overline n}$
will never stabilise close to unity when the loops fall inside the horizon.

Note that a loop with a very high initial
current will again `overshoots', but unlike in the case with friction here
it can actually end up oscillating faster than another one in the `current
branch' but with a smaller current. This is because now there is no friction
force that can damp this velocity overshoot.

It can therefore be seen that vortons can only
form during the friction-dominated epoch (as we expected), and also that
the earlier a loop is formed the larger will be the region of the
space of initial conditions that will originate them---because as we said
the effect of friction is to increase ${\overline n}$. Therefore, for cosmic
strings formed at the GUT phase transition, the most favourable case for
vorton formation is having the strings becoming superconducting at the GUT
scale as well. We will use this assumption in the remainder of the paper.

Finally, in figure \ref{chirsmall} we plot the more realistic case of the
evolution of GUT-scale loops having an initial string radius ten times smaller
than the horizon, and different initial ratios of energies in the current and
in the string---ranging from $10^{-3}$ to $2$.

Now the total radius only suffers a small decrease, except in the case where
one starts with ${\overline n}\sim1$, in which case velocity is so small that
friction does not significantly affect the loop. Note that as ${\overline n}$
approaches unity we have $v\propto t^{-1}$ as we predicted, although for loops
in the string branch there is an initial transient where $v\propto t^{-4}$.
Nevertheless, in the string branch loops do reach fairly high velocities during
their first few oscillations, so that once more the issue of whether or not
these become vortons is not entirely straightforward.

Also note that for loops of this size the amplitude of the energy
oscillations between the string and the current is negligibly small, except
for the short transient period (typically lasting less than one Hubble time)
for loops in the `string branch' with fairly
small currents. Clearly the relation between the initial conditions and the
values of ${\overline n}$ and $v$ needs to be looked at in more detail, and we
shall do that in the next section.

\section{Criteria for vorton formation}
\label{v-vt}
In the previous section we saw that the evolution of chiral superconducting
cosmic string loops depends sensitively on the conditions at formation. In
particular, one would need to know in which cases one ends up with a vorton.

Clearly, since we are not including radiative mechanisms at this stage, our
criterion should be that loops whose velocity is always small (in a sense
that will need to be made more precise)
will become vortons, while those who are relativistic
at some stage will suffer significant charge losses, so that their fate cannot
be clearly asserted until a rigourous quantum-mechanical treatment of these
processes is available.

Thus we will explore in more detail the phase space of possible
initial conditions in order to determine relevant properties of these loops.
Figure \ref{chmaxv} shows the maximum microscopic velocity
$v_{max}(\ell_i,t_i,{\overline n}_i)$ reached by GUT loops
formed at $t=t_c, 10\,t_c, 100\,t_c$ and $t_\star\sim855\,t_c$, respectively;
it is assumed that all such loops start their evolution with a negligibly small
velocity---a reasonable assumption, since the network dynamics is
friction-dominated until $t_\star$. In each case the horizontal axes correspond
to the initial value of ${\overline n}$ and to the base-ten logarithm
of the initial string
radius relative to the horizon; recall that
we only consider loops having initially most of
their energy in the string (in other words, loops in the string branch).
Note that the friction length-scale corresponds to
about $-1.5$ in the vertical axis on the first plot, and to $0$ on the
last (where it is equal to the horizon, by definition).

It can be seen that any loop initially
larger than the horizon will inevitably become relativistic.
This is essentially because expansion will (temporarily, at least) decrease
the fraction of the loop's energy in the current (and hence ${\overline n}$).
On the other hand, loops smaller than the friction length (and the horizon)
have essentially no mechanism that can change ${\overline n}$ (neglecting
radiation), so we will need fairly high initial currents in
order to get non-relativistic velocities.

Finally, for the case of loops being produced with sizes between the friction
length-scale and the horizon, which is of course the cosmologically relevant
case during the friction-dominated epoch \cite{ms}, friction will force
the loop to shrink (thereby increasing ${\overline n}$), while the effect of
the cosmological expansion will be small, so in order to have non-relativistic
velocities we are allowed to have smaller initial values of ${\overline n}$
than in the previous case.

From the analysis of figures \ref{chirrad}--\ref{chmaxv} one can see that we
need fairly high values of ${\overline n}$ when the loops reach the `free'
regime in order to have reasonable chances of producing  GUT vortons in the
`string branch'.
Now, according to Ref. \cite{vor}, the energy of a
superconducting loop configuration with radius R is approximately
\begin{equation}
E=2\pi\mu R+2\pi\Sigma\frac{W^2}{R}=E_{string}+E_{current}
\, ,\label{enerrda}
\end{equation}
where $W$ is the winding number and $2\pi\Sigma W\sim N_L^{1/2}$ is the
net particle number. The parameter $\Sigma$ is the result of an integral over
the string cross-section, it is a variable in general, but a constant in the
chiral case, and expected to be of the order of the inverse of a
coupling constant, $\Sigma\ge20$; we will in fact take $\Sigma=20$ unless
otherwise stated. This is minimised for a radius
\begin{equation}
\frac{R_v}{W}=\left(\frac{\Sigma}{\mu}\right)^{1/2}
\, ;\label{vorrad}
\end{equation}
this corresponds to a vorton state. As expected, this minimum value is
$E_v=2E_{string}$ and corresponds to ${\overline n}=1$.

Now, suppose that the energy of a given configuration is a little higher than
this minimum. That is, let $E=(1+x)E_v$. Then such a configuration will have
a radius
\begin{equation}
\frac{R_{string}}{R_v}=1+x\pm \left[(1+x)^2-1\right]^{1/2}
\, ;\label{pertradius}
\end{equation}
we will choose the plus sign since it corresponds to the `string branch'. Then
we can use (\ref{llsnbar2}) to find the corresponding value of ${\overline n}$:
\begin{equation}
{\overline n}=2\frac{1+x+\sqrt{x(2+x)}}{1+\left[1+x+\sqrt{x(2+x)}\right]^2}
\, .\label{pertnbar}
\end{equation}

These are useful expressions to introduce `phenomenological' criteria for
deciding which loop configurations will produce vortons. We note that these
should be established on the basis of more detailed numerical studies of
the microphysics of the currents; in particular, significant model-dependence
is of course expected.

As an example, if we take as a necessary condition for vorton formation
that the energy of a given configuration is at most $10\%$ higher than $E_v$,
we find that the value of ${\overline n}$ once the loop size becomes
smaller than the damping length should obey
\begin{equation}
{\overline n}_{free}\ge\frac{10}{11}\sim0.91
\, ,\label{cdvtn}
\end{equation}
or equivalently that the average fraction of the loop's energy in the current
must be
\begin{equation}
\langle \frac{E_{current}}{E_{total}}\rangle_T\ge0.45\, ;\label{cdvtr}
\end{equation}
Another (approximately) equivalent way of stating this is that a loop will not
form a vorton state if it exceeds some maximum velocity $v_{vor}$ above which
charge and current losses become effective. Note that a fast-moving loop will
tend to develop cusps at which such losses should be particularly
significant. Hence we require that $v_{max}<v_{vor}$ if a given loop is to
form a vorton. Of course $v_{vor}$ depends on $x$; for $x=0.1$, we have
\begin{equation}
v_{vor}\sim0.29\, . \label{cdvtv}
\end{equation}
Such a velocity limit is physically plausible, but a rigorous quantum
mechanical treatment will be required to obtain more precise values.
Note that the size of this vorton-forming region of parameter
space is maximal at $t_c$ and decreases with time, vanishing not later than
$t_\star$.

If we choose less stringent criteria, say $x=0.5$ or even
$x=1.0$, our bounds will respectively be
\begin{equation}
{\overline n}_{free}\ge\frac{2}{3}\, , \qquad v_{vor}\sim0.53\, ,\label{alt1}
\end{equation}
\begin{equation}
{\overline n}_{free}\ge\frac{1}{2}\, , \qquad v_{vor}\sim0.61\, ;\label{alt2}
\end{equation}
we will comment on the importance of the precise choice of $x$ in
section \ref{v-rs}.

Clearly, this only solves half of the problem---the other half
is determining what exactly are the initial conditions at the formation
of these loops, and in particular what are their currents. In other words,
we need to know whereabouts in figure \ref{chmaxv} do the loops form. This
is a non-trivial problem, but we will discuss
a simplified `toy model' for current evolution in the following section.

\section{Evolution of the Currents}
\label{v-cr}
Due to the strings' statistical nature, analytic evolution methods must be
`thermodynamic', that is one must describe the network by a small number
of macroscopic (or `averaged') quantities whose evolution equations
are derived from the microscopic string equations of motion. The first such
model providing a quantitative picture of the
complete evolution of a string network (and the corresponding loop
population) has been recently developed by the present
authors \cite{ms}, and we briefly summarise it here.

We start by defining our averaged quantities, the energy of a piece of
string,
\begin{equation}
E=\mu a(\tau)\int\epsilon d\sigma\, , \label{eee}
\end{equation}
($\epsilon$ being the coordinate energy per unit $\sigma$), and
the string RMS velocity,
defined by
\begin{equation}
v^2=\frac{\int{\dot{\bf x}}^2\epsilon d\sigma}{\int\epsilon d\sigma}
\, . \label{vv}
\end{equation}

Distinguishing between long (or `infinite') strings
and loops, and knowing that the former should be Brownian
we can define the long-string correlation length as
$\rho_{\infty}\equiv\mu/L^2$ (see \cite{ms} for an extensive
discussion of these quantities, and others to be introduced below).
A phenomenological term must then be included for the interchange
of energy between long strings and loops. A `loop chopping
efficiency' parameter, expected to be slightly smaller than unity, is
introduced to characterise loop production
\begin{equation}
\left(\frac{d\rho_{\infty}}{dt}\right)_{\rm to\ loops}=
{\tilde c}v_\infty\frac{\rho_{\infty}}{L}
\, . \label{rtl}
\end{equation}

One can then derive the evolution equation for the
correlation length $L$ \cite{ms}, which has the form
\begin{equation}
2 \frac{dL}{dt}=2HL(1+v_\infty^2)+
v_\infty^2\frac{L}{\ell_f}+{\tilde c}v_\infty \, ; \label{evl}
\end{equation}
we point out again that the `friction lengthscale' $\ell_{\rm f}$ will
in general be that due to Everett scattering.

One can also derive an evolution equation for the long string
velocity with only a little more than Newton's second law
\begin{equation}
\frac{dv}{dt}=\left(1-v^2\right)\left[\frac{k}{L}-v\left(2H+
\frac{1}{\ell_f}\right)\right] \, ;
\label{evv}
\end{equation}
here $k$ is another phenomenological parameter that is equal to unity
during the friction-dominated epoch and of order unity later \cite{ms}.

Finally, a careful analysis of the loop production mechanism leads to an
expression for the energy density in loops. The idea is that at a given time
one looks back at all the loops that have formed (and still have not decayed),
finds their present lengths and then adds them together. Distinguishing
between `dynamical' and `primordial' (that is, Vachaspati-Vilenkin) loops,
we have
\begin{equation}
\rho_o(t)=\int_{t_c}^tn_{dyn}(t,t')\ell(t,t')dt'+\int_{L_c}^{L_{cut}}
n_{pri}(\ell',t)\ell_{pri}(\ell',t)d\ell'\, . \label{oldrho}
\end{equation}
Above $\ell(t,t')$ is the length at time $t$ of a loop produced at time $t'$
(this will vanish if the loop has decayed), while
$n_{dyn}(t,t')a^3(t)=n_{loop}(t')a^3(t')$, where
\begin{equation}
n_{loop}(t)=g\mu{\tilde c}\frac{v_\infty}{\alpha L^4}\, \label{loppr}
\end{equation}
is the number of loops produced per unit time per unit volume.
The factor $g\sim1/\sqrt{2}$ accounts for the fact that not all of the energy
lost by the long-string network ends up in the loops---part of it is lost by
velocity redshift. We are assuming
that loops produced at time $t$ have an initial length
$\ell(t)=\alpha(t)L(t)$---in other words, that loop production is
`monochromatic' (see \cite{ms} for a discussion of this point). Similarly, for
the Vachaspati-Vilenkin loops $\ell_{pri}(\ell',t)$ is the length at time $t$
of a loop formed with length $\ell'$, while
$n_{pri}(\ell',t)a^3(t)=n_{loop}(\ell')a^3(t_c)$, where $n_{loop}(\ell')$
is the well-known Vachaspati-Vilenkin loop distribution.

The above quantities are sufficient to quantitatively
describe the large-scale characteristics of a cosmic string network.
We will describe the evolution of the currents by a recently introduced
toy model \cite{mss}, which we now discuss in more detail.

Our analysis will be based on the assumption that there is a `superconducting
correlation length', denoted $\xi$, which measures the scale over which
one has coherent current and charge densities on the strings. Associated with
this we can define $N$ to be the number of uncorrelated current regions
(in the long-string network) in a co-moving volume $V=a^3L_0^3$ as follows
\begin{equation}
N\equiv\frac{L_\infty}{\xi}=\frac{V}{\xi L^2}\, ,\label{defni}
\end{equation}
where $L_\infty$ is the total long string length in the co-moving volume.

\begin{figure}
\vbox{\centerline{
\epsfxsize=1.0\hsize\epsfbox{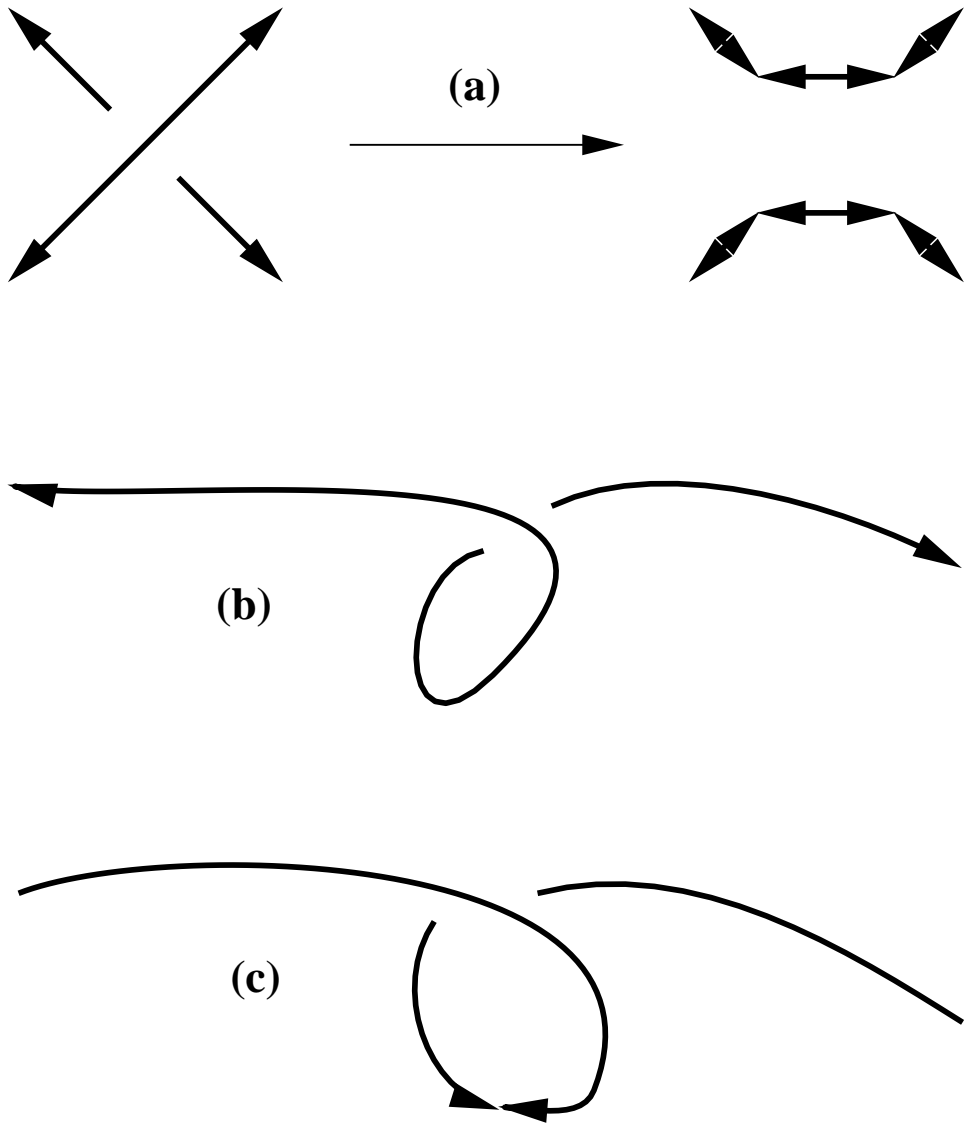}}
\vskip.4in}
\caption{Some relevant inter-commuting configurations. The arrows mark the
limits of regions with correlated currents. Plot (a) shows a typical
inter-commuting creating four new current regions, while (b-c) show than
on scales smaller than the current correlation length loop production may (c)
or may not (b) remove current regions from the long-string network.}
\label{bonec}
\end{figure}

Now, $\xi$ and $N$ will obviously change in the course of the evolution of the
string network, and we can immediately identify four possible sources of
change---expansion, inter-commuting, loop production and internal dynamics
on the string worldsheet. We now consider each one of them.
Firstly, we expect that in a co-moving volume the number of uncorrelated
regions will not be affected by expansion, so
\begin{equation}
\left(\frac{dN}{dt}\right)_{expansion}=0\, .\label{expterm}
\end{equation}
Now consider the effect of inter-commutings (whether or not a
loop is produced).
Laguna and Matzner \cite{laguna} have numerically shown that whenever two
current-carrying strings cross, they inter-commute and a region of
intermediate current is created. This means that inter-commutings
will in general create four new regions (see figure \ref{bonec} (a)).
Since, according to our analytic evolution model the inter-commuting rate is
\begin{equation}
\left(\frac{dn}{dt}\right)_{intercommuting}=\frac{1}{2}
\frac{v_\infty}{\alpha}\frac{V}{L^4}\, , \label{itcrate}
\end{equation}
we immediately obtain the following effect on $N$
\begin{equation}
\left(\frac{dN}{dt}\right)_{intercommuting}=2\frac{v_\infty}{\alpha}
\frac{V}{L^4}\, ; \label{incterm}
\end{equation}
again this  assumes that loops have a size $\ell(t)=\alpha(t) L(t)$ at
formation,
and that once the long-string network reaches the linear scaling regime we
have $\alpha(t)=\alpha_{sc}=const.$ (see \cite{ms}).

However, an important correction is necessary to account for the fact that
when regions with size of order $\xi$ or smaller self-intersect it is possible
(see figure \ref{bonec} (b-c)) that no new regions are produced. Thus we
must multiply (\ref{incterm}) by a correction factor
\begin{eqnarray}
F_1\left(\frac{\ell}{\xi}\right)= 1 & , &
\mbox{$\frac{\ell}{\xi}>1$} \\
F_1\left(\frac{\ell}{\xi}\right)=
\alpha\left[1-2{\tilde c}\left(1-\frac{\ell}{\xi}\right)\right]+
(1-\alpha)\frac{\ell}{\xi} & , &
\mbox{$\frac{\ell}{\xi}\le1$}\, .
\label{gans}
\end{eqnarray}
The slightly complicated behaviour of $F_1$ is nevertheless easy to understand.
The point is that numerical simulations show that there are two types of
inter-commutings. Firstly, `large-scale' ones always occur at a scale $L$; a
fraction $\alpha$ of the inter-commutings should be of this type. If this
happens between two long-strings (that is, no loop is produced) we always
expect to create new regions, since there is no reason for currents
in different `infinite' stings to be correlated. On the other hand,
if what we have is a long
string self-intersecting to produce a loop of size smaller than $\xi$ (a
fraction $2{\tilde c}$ of these inter-commutings should produce loops),
then we might not form new regions---for each
length, the fraction of these self-intersections that produce new regions is
essentially given by the ratio of the size of the region and the
superconducting correlation length. The remainder of the inter-commutings
are associated with the presence of small-scale structure on the strings, and
occur by repeated self-intersections of a given string so the $\ell/\xi$
cutoff always applies. Notice that the second term vanishes if $\alpha=1$
(as it should) but it rapidly becomes dominant as $\alpha$ starts deviating
from unity. Also note that the overall inter-commuting effect is approximately
$\alpha$-independent (more on this below).

Of course, when the inter-commuting
does produce a loop, the regions in the corresponding segment are removed
from the network, together with one of the newly created
`intermediate' regions, and we similarly have
\begin{equation}
\left(\frac{dN}{dt}\right)_{loops}=-\left(\frac{\ell}{\xi}+2F_2\right)
{\tilde c}\frac{v_\infty}{\alpha}\frac{V}{L^4}\, ,
\label{lppterm}
\end{equation}
where the analogous correction factor $F_2$ is of the form
\begin{eqnarray}
F_2\left(\frac{\ell}{\xi}\right)=1 & , &
\mbox{$\frac{\ell}{\xi}>1$} \\
F_2\left(\frac{\ell}{\xi}\right)=\frac{\ell}{\xi} & , &
\mbox{$\frac{\ell}{\xi}\le1$} \, .
\label{gans2}
\end{eqnarray}
Note that string length is always removed from the long string network when
loops form, regardless of whether
or not current regions are. This is in fact the main effect of loop
production, as can be seen by noting that (\ref{lppterm}) is approximately
independent of the parameter characterising the loop size, $\alpha$.
In the friction-dominated regime, $\alpha$ is of order unity, and when it
becomes much smaller (in the free regime) the $\alpha$-dependencies in
the numerator and in the denominator cancel out. One can readily see that
this is physically plausible: when $\alpha\sim1$ (in the friction-dominated
epoch) few loops are produced, but each one of them removes a significant
number of regions; on the other hand, when $\alpha$ is small, many more
loops are produced, but only a few of them will remove regions.

Finally, there is the dynamic term. When regions with opposite currents
inter-commute, new charged regions are created, setting up alternate currents.
One expects electromagnetic processes to make these currents die down,
so that the charged region will eventually equilibrate with its
neighbours. The simulations of Laguna and Matzner \cite{laguna}
provide qualitative support for this intuitive picture. Clearly, this indicates
that some kind of `equilibration' process is effectively acting between
neighbouring current regions, which will counteract the creation of new regions
by inter-commuting. While it is beyond our means to derive an `equilibration
term' from first principles we will, as a first approximation, introduce a
phenomenological term. We will model this current decay by assuming that after
each Hubble time, a fraction $f$ of the $N$ regions existing at its start will
have equilibrated with one of its neighbours,
\begin{equation}
\left(\frac{dN}{dt}\right)_{dynamics}=-fHN\, ;
\label{eqlterm}
\end{equation}
note that new regions are obviously created by inter-commuting during the
Hubble time in question, so that $f$ can be larger than unity.
Alternatively we can say that for a given $f$, the number of regions in a
given volume at a time $t$ will have disappeared due to equilibration at a
time $t+(fH)^{-1}$.
We therefore obtain the following evolution equation for $N$
\begin{equation}
\frac{dN}{dt}=G\left(\frac{\ell}{\xi}\right)
\frac{v_\infty}{\alpha}\frac{V}{L^4}-fHN\, ,
\label{generalnn}
\end{equation}
where we have re-defined the correction factor
\begin{eqnarray}
G\left(\frac{\ell}{\xi}\right)=
2-{\tilde c}\left(\frac{\ell}{\xi}+2\right) & , &
\mbox{$\frac{\ell}{\xi}>1$} \\
G\left(\frac{\ell}{\xi}\right)=
2(1-2{\tilde c})\alpha+(2-3{\tilde c}-2\alpha+4\alpha{\tilde c})
\frac{\ell}{\xi} & , & \mbox{$\frac{\ell}{\xi}\le1$} \, .
\label{gansg}
\end{eqnarray}
Note that when $\ell\gg\xi$ the net effect of inter-commuting and loop
production is to remove uncorrelated regions (because each loop formed
removes a large number of them); otherwise, the net effect is to create new
regions.

However, for what follows it is convenient to re-write it in two alternative
forms. Firstly, we can define $N_L$ to be the number of uncorrelated
current regions per long-string correlation length,
\begin{equation}
N_L\equiv\frac{L}{\xi}=N\frac{L}{L_\infty}\, ;\label{defnl}
\end{equation}
this is useful because, as was first pointed out by Davis and
Shellard \cite{vor}, we expect the net charge of a superconducting loop to be
given by
\begin{equation}
Q\sim eN_L^{1/2}\, .\label{ccvor}
\end{equation}
In terms of $N_L$, (\ref{generalnn}) has the form
\begin{equation}
\frac{dN_L}{dt}=(3v_\infty^2-f)HN_L+\frac{3}{2}
\frac{v_\infty^2}{\ell_{\rm f}}N_L+\left(\frac{1}{\alpha}G(\alpha N_L)+
\frac{3}{2}{\tilde c}N_L\right)\frac{v_\infty}{L}\, ;
\label{generalnl}
\end{equation}
note that to obtain this one needs to substitute the evolution equation for the
long-string correlation length $L$ (\ref{evl}), and that one can
equivalently define $G$ as
\begin{eqnarray}
G\left(\alpha N_L\right)=
2-{\tilde c}\left(\alpha N_L+2\right) & , &
\mbox{$\alpha N_L>1$} \\
G\left(\alpha N_L\right)=2(1-2{\tilde c})\alpha+(2-3{\tilde c}
-2\alpha+4\alpha{\tilde c})\alpha N_L & , &
\mbox{$\alpha N_L\le1$} \, .
\label{newgans}
\end{eqnarray}

Yet another useful form follows from defining $N_H$ to be the number of
uncorrelated current regions in one Hubble volume,
\begin{equation}
N_H\equiv\frac{L_H}{\xi}=N_L\frac{d_H^3}{L^3}\, ;\label{defnh}
\end{equation}
in this case we have
\begin{equation}
\frac{dN_H}{dt}=(3-f)HN_H+G\left(\alpha N_L\right)\frac{v_\infty}{\alpha}
\frac{d^3_H}{L^4}\, .
\label{generalnh}
\end{equation}

\section{The Importance of Equilibration}
\label{v-ff}
Now the question is, of course, what is $f$?
From a more intuitive point of view, an equivalent question is the following:
given a particular piece of string with a given current, is it more likely to
disappear from the network by this equilibration mechanism or by being
incorporated in a loop?
Even though a precise answer
can probably only be given by means of a numerical simulation, some
very simple physical arguments can be used to restrict it. We should
point out, however, that many of the results of the following sections do not
depend crucially on the value of $f$.

Firstly,
correlations cannot obviously be established faster than the speed
of light (that is, we must have $\xi\le t$), so that we should impose that
\begin{equation}
\left(\frac{dN_H}{dt}\right)_{N_H=L_H/t}\ge0\, ; \label{minnh}
\end{equation}
this leads to an upper bound on $f$, which we can write, defining $L=\gamma t$,
as
\begin{equation}
f_{max}({\tilde c},\gamma,v_\infty)=3+4(1-2{\tilde c})
\frac{v_\infty}{\gamma^2}\, .\label{maxff}
\end{equation}
(In this section we will concentrate on the bounds on $f$ in the radiation
epoch---analogous results can obtained for the matter epoch.)
We explicitly write the dependencies of $f_{max}$ to emphasise that this is
the maximum value of $f$ which satisfies (\ref{minnh}) for a given set of
properties of the cosmic string network.

On the other hand, if an equilibration mechanism such as that modelled by
(\ref{eqlterm}) exists \cite{laguna}, it is reasonable to assume that
it will prevent $N_H$ from growing without limit---possibly through a
backreaction mechanism as in the case of gravitational radiation for wiggly
Goto-Nambu strings---and eventually it will make it become constant (meaning
that $\xi$ is scaling linearly). In other words, we can assume that there
should be a large $N_H^\star$ (which we need not specify) such that
\begin{equation}
\left(\frac{dN_H}{dt}\right)_{N_H^\star}\le0\, ;\label{maxnh}
\end{equation}
we can therefore find a lower bound on $f$ which satisfies this,
\begin{equation}
f_{min}({\tilde c},\gamma,v_\infty)=3-2{\tilde c}\frac{v_\infty}{\gamma}
\, .\label{minff}
\end{equation}
Again this varies as the network evolves.
Note that the crucial point about this construction is that (\ref{eqlterm})
depends linearly on $N_H$.

\begin{figure}
\vbox{\centerline{
\epsfxsize=0.8\hsize\epsfbox{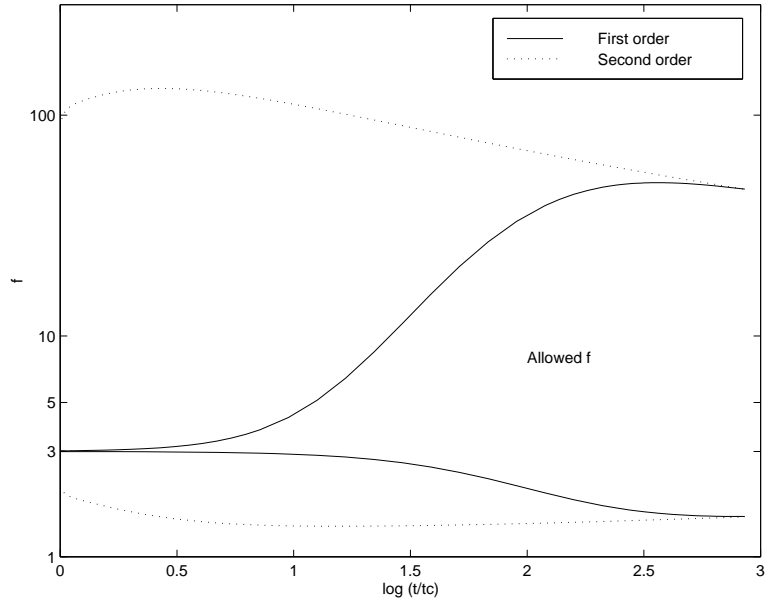}}
\vskip.4in}
\caption{The evolution of $f_{min}$ (lower pair of curves) and $f_{max}$
(upper pair of curves) for first order (solid lines)
and second order (dotted lines) string-forming phase transitions. Time is in
orders of magnitude from the epoch of string formation.}
\label{fbds}
\end{figure}

Using the quantitative evolution model of the present authors we have plotted
$f_{max}$ and $f_{min}$ during the friction-dominated epoch with initial
conditions typical of first and second
order phase transitions, in figure \ref{fbds}. These plots are fairly easy to
interpret. Perhaps the most surprising result is the large values of
$f_{max}$ allowed when the long-string correlation length is well below the
horizon. This is because in this case
the loops chopped off by the network are small, so that each one of them
removes relatively few current regions---one could therefore have an extremely
efficient equilibration mechanism and still obey the constraint (\ref{minnh}).
Hence we can see from figure \ref{fbds} that if our toy model, and in
particular the {\em ansatz} (\ref{eqlterm}) is valid the constraints on $f$ are
much stronger for a first-order phase transition.
One can also see that $f=3$ is
the only value that is acceptable at all times, regardless of the initial
condition. However, it is at present unclear if there is something `special'
about this value. A numerical simulation is presumably the only way to
clarify this issue.

Note that any value $f_{min}<f<f_{max}$ once the network reaches the linear
scaling regime
\begin{equation}
1.88\sim3-\frac{2{\tilde c}}{k+{\tilde c}}<f<3+\frac{4(1-
2{\tilde c})}{k^{1/2}(k+{\tilde c})^{3/2}}\sim22.4\, ,\label{ffscal}
\end{equation}
leads to a constant value of $N_L$ and that this corresponds
to $\xi$ scaling as the long-string correlation length $L$. Different values
of $f$ lead to different scaling values of $N_L$ (with larger $f$'s
corresponding to smaller $N_L$'s as expected) at least in some region
of the space of initial conditions, but for any $f\neq3$ one can think of
some set of physically viable initial conditions for which either
causality would be violated at some stage of the evolution or the number of
uncorrelated regions would grow without bound. The scaling value of $N_L$
can be written in terms of the properties of the string network as
\begin{equation}
\alpha_{sc} N_L=\frac{4(1-{\tilde c})}{(k+{\tilde c})f-{\tilde c}-3k}
\, ;\label{scalnsl}
\end{equation}
one can see that in this regime the $f$ dependence is rather weak,
unless $f$ is just above $f_{min}$.

We emphasise that while the $f_{max}$ bound is unavoidable (being a consequence
of causality), $f_{min}$ is less robust and could well be disproved by a
detailed numerical study. Therefore, in what follows we will discuss
two cases, $f=0$ and $f=3$ which should represent the scenarios of
ineffective (or non-existent) and effective equilibration.

For a given $f$, we can now solve (\ref{generalnl}) numerically, coupled
with the evolution equations for the long-string correlation length and
average velocity (see \cite{ms}). This therefore allows us to know the size
of the loops formed by the network at each time and (through (\ref{ccvor})) the
initial current they will carry. On is then in a position of applying the
criteria established in section \ref{v-vt} in order to decide whether or not
each loop will form a vorton.

We should also say at this stage that once the network leaves the
friction-dominated regime and strings become relativistic other mechanisms
(notably radiation) can cause charge losses in the long strings (as well as
in loops). Hence our toy model can at best provide order-of-magnitude
estimates in this regime. On the other hand, we expect it to be quite accurate
(pending a more detailed numerical study) in the friction-dominated
epoch---which is of course relevant for vorton formation.

\section{GUT-scale Analysis}
\label{v-rs}
\begin{figure}
\vbox{\centerline{
\epsfxsize=0.6\hsize\epsfbox{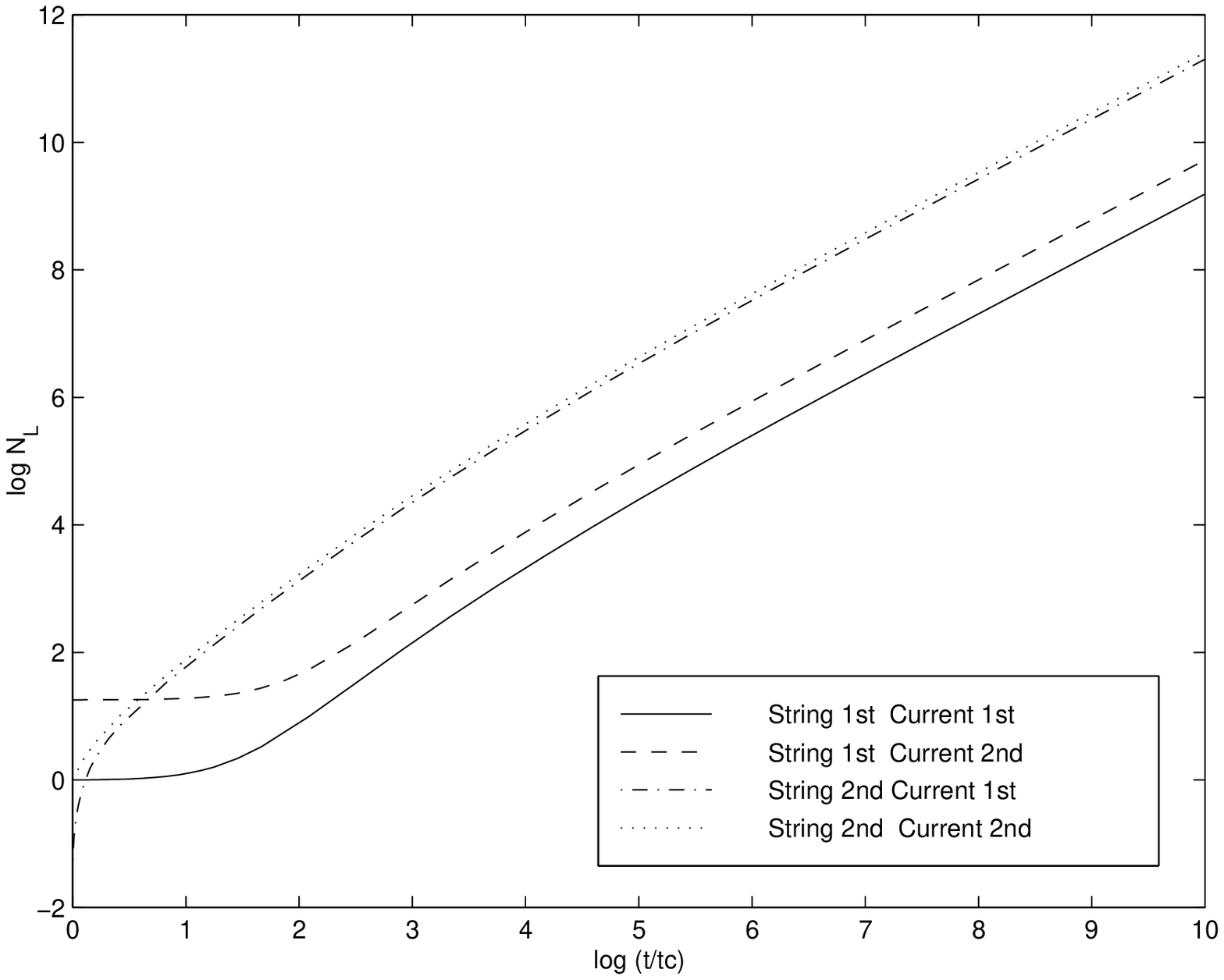}}
\vskip.4in}
\vbox{\centerline{
\epsfxsize=0.6\hsize\epsfbox{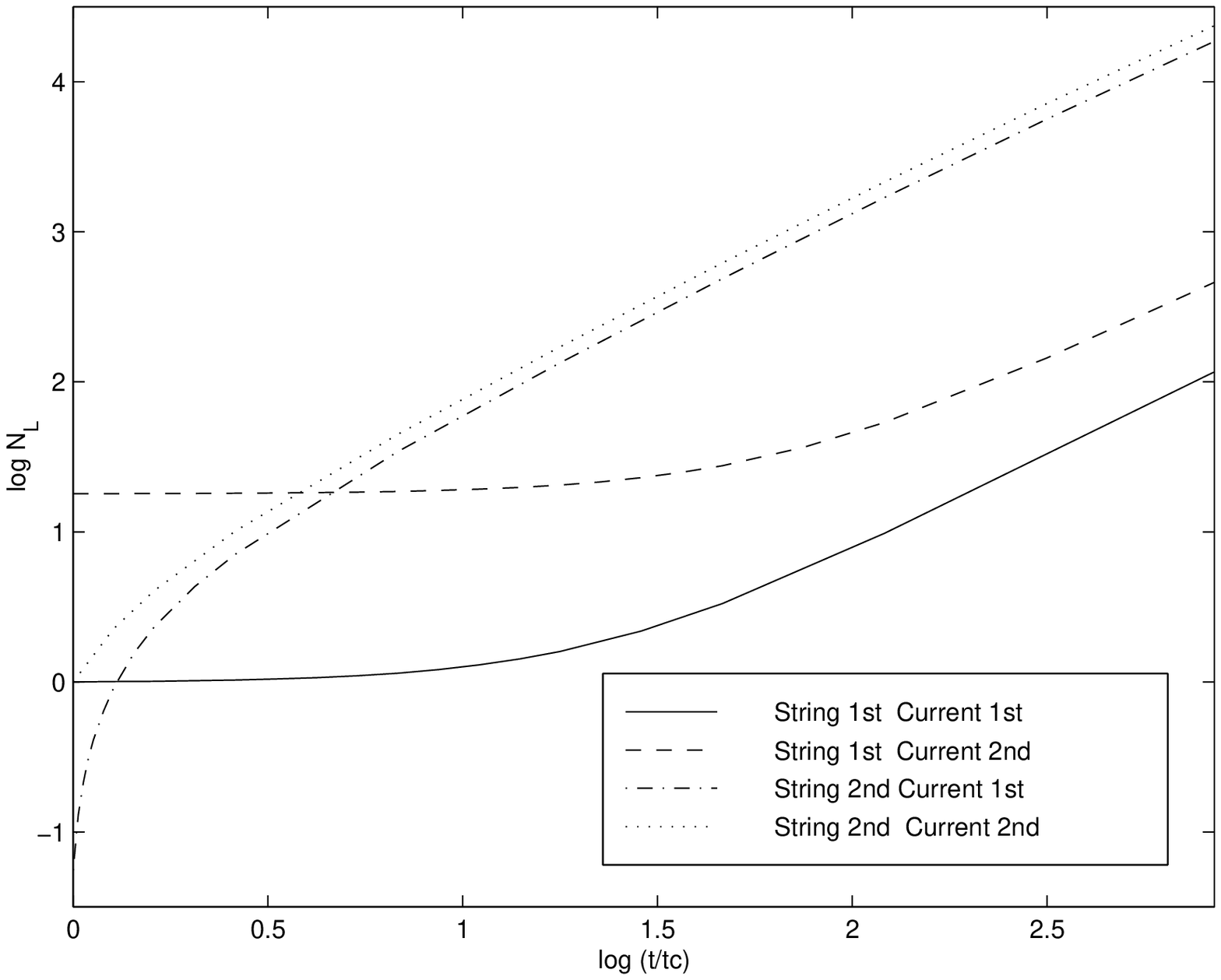}}
\vskip.4in}
\caption{The evolution of the number of uncorrelated current regions per
long-string correlation length, $N_L$, for the case $f=0$ (the bottom
plot is a friction-dominated epoch close-up of the top one) assuming that
the orders of the string-forming
and superconducting phase transitions are respectively:
1st \protect\& 1st (solid lines), 1st \protect\& 2nd (dashed),
2nd \protect\& 1st (dash-dotted) and 2nd \protect\& 2nd (dotted). Time is in
orders of magnitude from the epoch of string formation.}
\label{evn_l0}
\end{figure}

\begin{figure}
\vbox{\centerline{
\epsfxsize=0.6\hsize\epsfbox{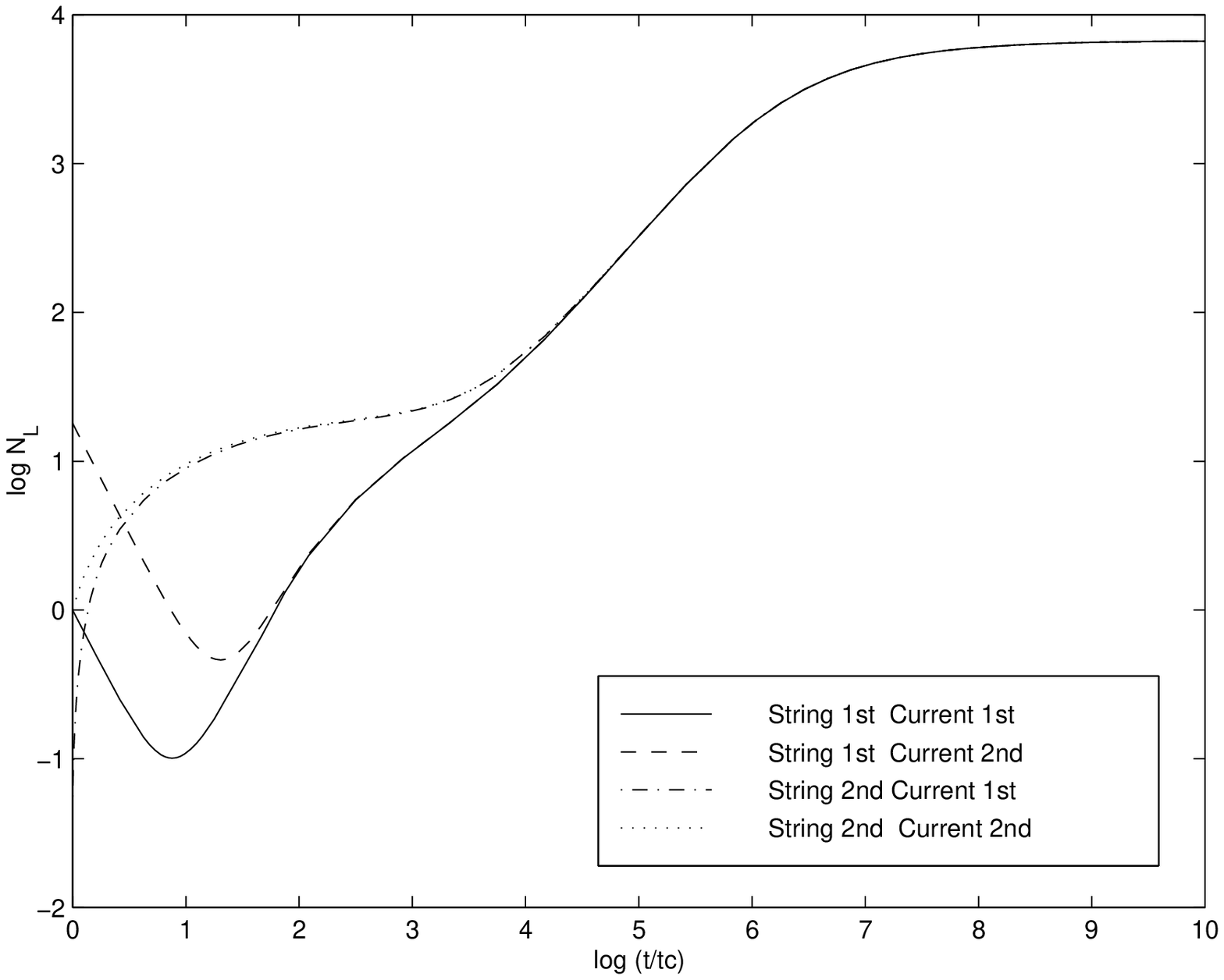}}
\vskip.4in}
\vbox{\centerline{
\epsfxsize=0.6\hsize\epsfbox{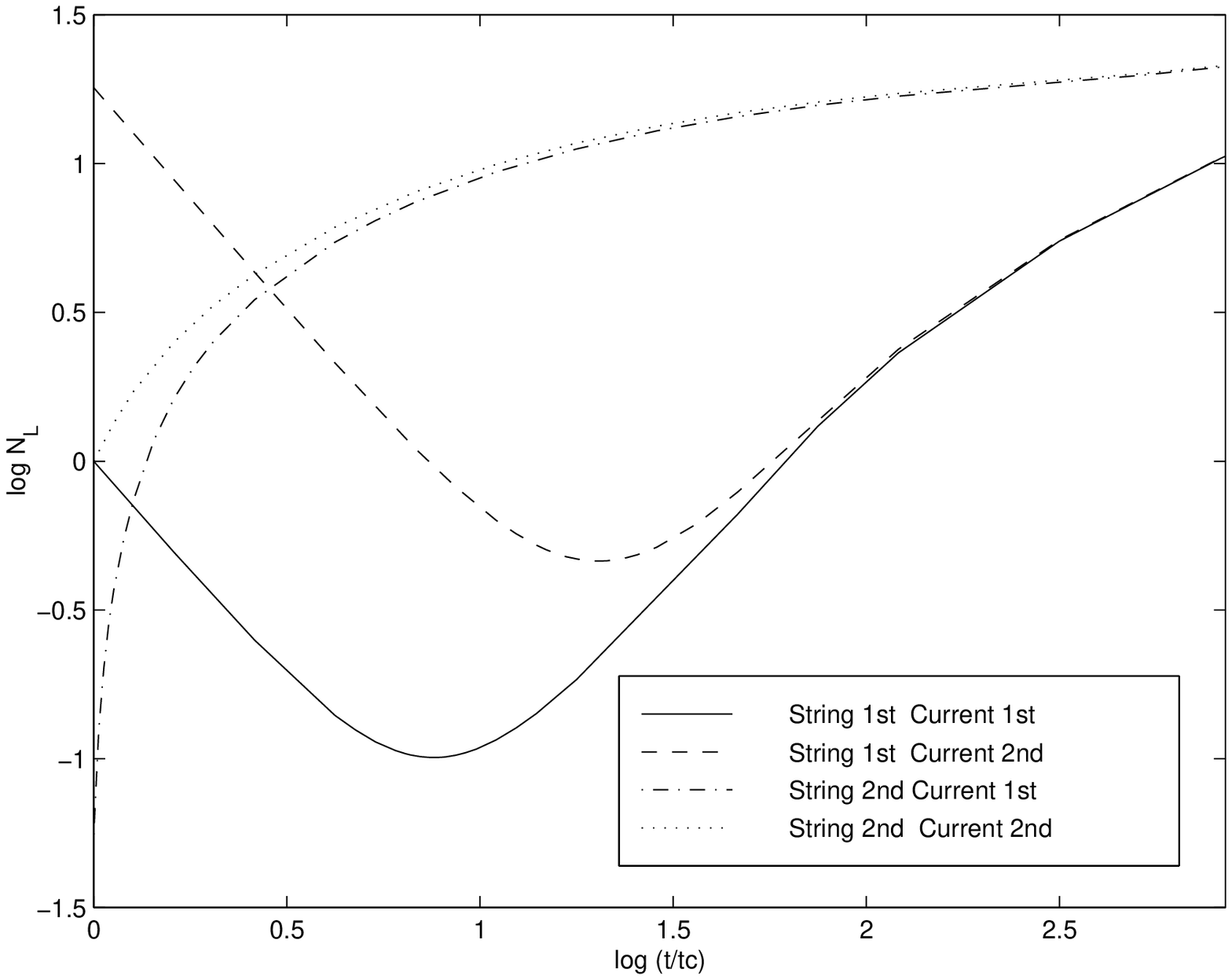}}
\vskip.4in}
\caption{The evolution of the number of uncorrelated current regions per
long-string correlation length, $N_L$, for the case $f=3$ (the bottom
plot is a friction-dominated epoch close-up of the top one) assuming that
the orders of the string-forming
and superconducting phase transitions are respectively:
1st \protect\& 1st (solid lines), 1st \protect\& 2nd (dashed),
2nd \protect\& 1st (dash-dotted) and 2nd \protect\& 2nd (dotted). Time is in
orders of magnitude from the epoch of string formation.}
\label{evn_l3}
\end{figure}

In figures \ref{evn_l0}-\ref{evn_l3} we plot the result of the numerical
integration of (\ref{generalnl}), for initial conditions
representative of string-forming and superconducting phase transitions of
first and second order, for the cases $f=0$ and $f=3$. We are assuming that
these occur at around the same
(GUT) energy scale since, as we have shown in section \ref{v-vt}, this is
most favourable situation for vorton formation. It was also assumed that
the value of $\alpha$ in the linear scaling regime is $\alpha_{sc}\sim10^{-3}$
(see ref. \cite{ms}).

The differences between the two cases are considerable. Firstly, if there is no
equilibration mechanism ($f=0$, see figure \ref{evn_l0}), the number of
uncorrelated regions per long-string correlation length $N_L$, never
decreases. In this case there are simple scaling laws for $N_L$ and $\xi$.
One finds that $\xi$ is conformally stretched during the
stretching regime (just like the long-string correlation length,
$L\propto t^{1/2}$), and
so $N_L$ is approximately constant. However, as inter-commutings start
creating new regions $N_L$ begins to increase, and it grows as $t^{3/2}$ during
the Kibble regime (where $L\propto t^{5/4}$, so $\xi\propto t^{-1/4}$).
Finally, once the network reaches the linear scaling regime, $L\propto t$,
the number of uncorrelated current regions grows as $N_L\propto t$, which
corresponds to $\xi\propto const$. As expected, in this case the network
keeps a `memory' of its initial conditions.

On the other hand, if there is an efficient enough
equilibration mechanism (see figure \ref{evn_l3} for the case $f=3$) then $N_L$
decreases while the network is being conformally stretched. In the Kibble
regime, the increased number of inter-commutings again drives $N_L$ up,
and after $\alpha$ has evolved into its linear regime value, $\xi$ itself
reaches a scaling value and hence $N_L$ becomes a constant. In the
intermediate case of a small but non-zero $f$, $N_L$ decreases during the
stretching regime but grows without limit afterwards, and the precise values
of the scaling laws depend on $f$. Also, the network will preserve a `memory'
of the order of the string-forming phase transition, but not of the order
of the superconducting one.

This therefore solves the other half of our problem. Knowing the loop size
at formation at all times \cite{ms} at the typical current that each loop
carries at that epoch (from the above toy model) one can then apply some
criterion (possibly of the type discussed in  section \ref{v-vt}) to
decide which loops have a reasonable possibility of becoming vortons.

Our quantitative string evolution model \cite{ms} allows us to determine
the size of the loops formed at each epoch, $\ell(t)=\alpha(t)L(t)$. On the
other hand, according to (\ref{llsnbar2}), to find the initial ${\overline n}$
we need to know the ratio of the energies in the string and in the current.
Now, the energy of a superconducting loop configuration with a radius R is
given by (\ref{enerrda}), so after some algebra we find
\begin{equation}
\frac{E_{current}}{E_{string}}=\frac{16\pi^3{\cal N}}{45 \Sigma}
\frac{G\mu N_L}{\alpha^2 \gamma^2 x^2}\, ,\label{detnbar}
\end{equation}
where $L=\gamma t$, $t=xt_c$ and ${\cal N}$ is the number of effectively
massless degrees of freedom. Note that the minimum value of $\gamma$ is
of the order of $(G\mu)^{1/2}$ (but slightly larger---see \cite{ms}),
so the crucial factor
in this equation, and hence for vorton formation, is how much $N_L$ can
grow. This alone tells us that the higher the energy scale at which the
string network forms, the less likely it is to produce vortons, since it
will be friction-dominated (and hence non-relativistic) for a shorter period
of time. In order to make vortons, loops should be formed with a high enough
$N_L$ to allow them to remain non-relativistic thereafter---otherwise, they
will eventually become relativistic and hence liable to charge losses. As we
already pointed out, making the strings become superconducting sometime after
they form does not  help---it merely reduces the time available to build up
charges and currents.

\begin{figure}
\vbox{\centerline{
\epsfxsize=0.6\hsize\epsfbox{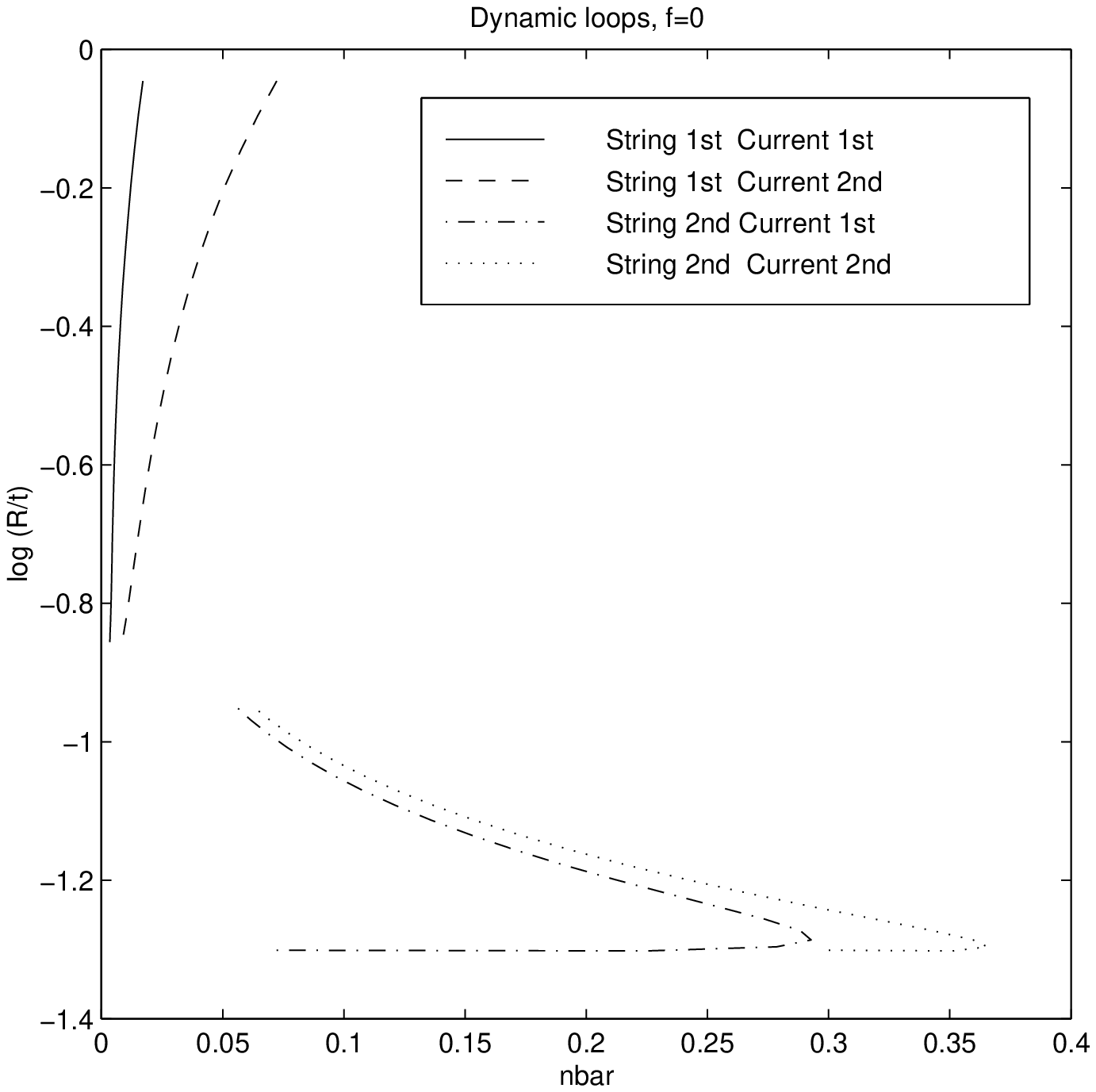}}
\vskip.4in}
\vbox{\centerline{
\epsfxsize=0.6\hsize\epsfbox{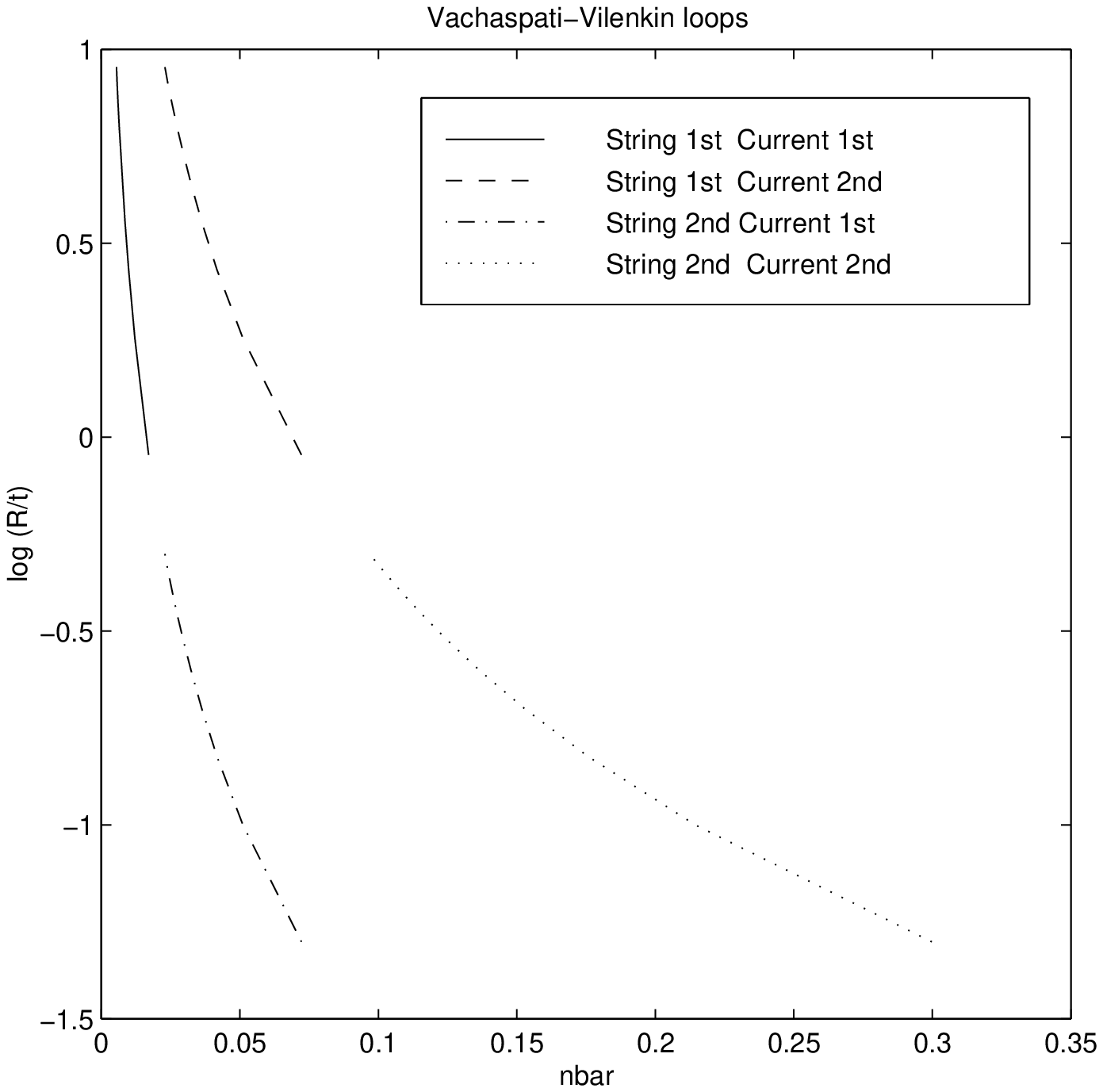}}
\vskip.4in}
\caption{The initial conditions for loop formation in
${\protect\overline n}$--$R$ space, for the case $f=0$, assuming that the
orders of the string-forming and superconducting phase
transitions are respectively:
1st \protect\& 1st (solid lines), 1st \protect\& 2nd (dashed),
2nd \protect\& 1st (dash-dotted) and 2nd \protect\& 2nd (dotted).
The top plot corresponds to dynamic loops formed between $t_c$ and $100\,t_c$
(in the first two curves loops formed at $t_c$ are at large $R$; in the later
two they are at small $R$).
The bottom corresponds to `primordial' Vachaspati-Vilenkin loops formed at
$t_c$ and having lengths between $L_c$ and $10\,L_c$.}
\label{paths0}
\end{figure}

\begin{figure}
\vbox{\centerline{
\epsfxsize=0.6\hsize\epsfbox{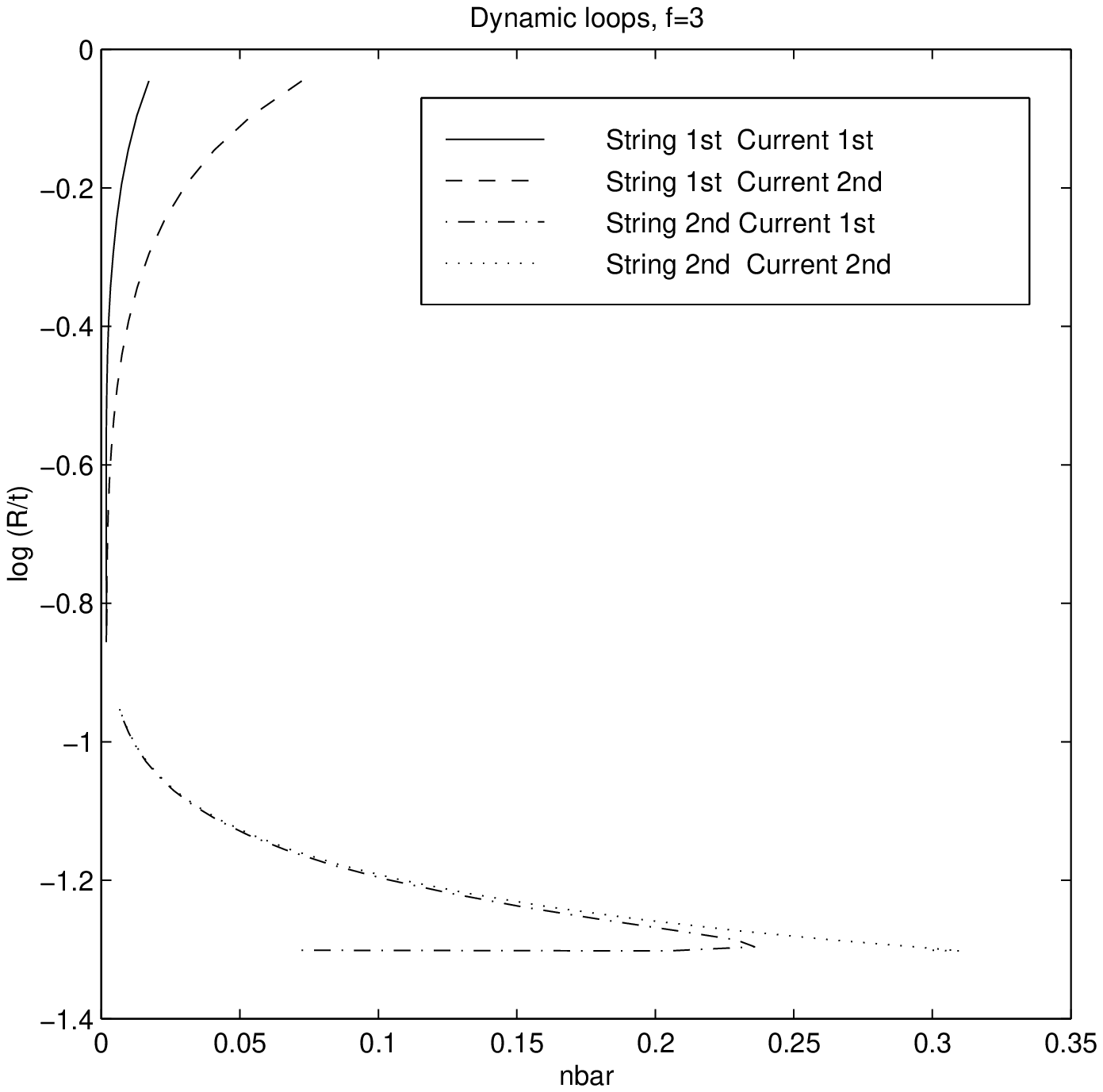}}
\vskip.4in}
\vbox{\centerline{
\epsfxsize=0.6\hsize\epsfbox{fig11b12b.eps}}
\vskip.4in}
\caption{The initial conditions for loop formation in
${\protect\overline n}$--$R$ space, for the case $f=3$, assuming that the
orders of the string-forming and superconducting phase
transitions are respectively:
1st \protect\& 1st (solid lines), 1st \protect\& 2nd (dashed),
2nd \protect\& 1st (dash-dotted) and 2nd \protect\& 2nd (dotted).
The top plot corresponds to dynamic loops formed between $t_c$ and $100\,t_c$
(in the first two curves loops formed at $t_c$ are at large $R$; in the later
two they are at small $R$).
The bottom corresponds to `primordial' Vachaspati-Vilenkin loops formed at
$t_c$ and having lengths between $L_c$ and $10\,L_c$.}
\label{paths3}
\end{figure}

Contrary to current belief (which is based on rather more
qualitative estimates) we do not  expect any vortons to be produced
by GUT-scale cosmic string networks.
In figures \ref{paths0}-\ref{paths3} we plot the paths of initial conditions
in ${\overline n}$--$R$ space for  dynamic and Vachaspati-Vilenkin loops
\cite{vacha} formed during the friction-dominated epoch in the cases
$f=0$ and $f=3$. For the `dynamic' loops, we only plot
loops formed until $100\,t_c$ (notice that ${\overline n}$ decreases after this
epoch). We consider initial
conditions for the string network that are characteristic of first- and
second-order string-forming and superconducting phase transitions.
Note that the difference in the initial ${\overline n}$'s between the two
cases is smaller than the difference between the corresponding $N_L$'s;
this is because ${\overline n}$ is approximately proportional to $N_L^{1/2}$.

One can see that, even if we choose the less stringent of our three
suggested criteria, calling a vorton any loop configuration with
an energy up to twice the minimum value (that is, $x=1$) we still get
no GUT vortons.
In fact, one would need to choose a limiting velocity $v_{vor}\sim0.7$ for GUT
vorton production to occur in this model---and even so, only in the case when
equilibration is efficient and the string-forming phase transition is of
second order. However, we should emphasise the issues of the precise vorton
formation criterion, as well as that of the value of the `equilibration
parameter' $f$, can only be settled by means of more detailed numerical
studies of the microphysics of these loop configurations.

This is an appropriate point at which to add a cautionary
note about the quantum-mechanical stability of vortons.
This is a rather involved and model-dependent question which
has been briefly discussed in ref. \cite{vor}. The vorton gains
additional stability because its charge carriers must
tunnel off the string by taking both charge and angular
momentum; the larger a vorton is, the more stable it is.
With electromagnetic fields present, pair creation provides an alternative
decay mechanism, but for a chiral vorton with null fields ($E^2=B^2$) near
the string, this mechanism is strongly suppressed. As the chiral state
tends to be an attractor for a wide range of initial conditions \cite{davsh}
this again encourages us to believe that vortons should generically be
quantum mechanically stable. Nevertheless, one can make special parameter
choices for which vorton lifetimes are very brief, notably when the string
and current-forming phase transitions are widely separated in energy scale.
This subject clearly deserves a more thorough investigation.

\section{Calculating vorton densities}
\label{v-den}
Vorton densities can be calculated using a fairly straightforward modification
to the method developed in \cite{ms} and summarised above.
We now have
\begin{eqnarray}
\rho_v(t)&=&\int_{t_c}^{t}W_1(t')n_{dyn}(t,t')\ell(t,t')dt'\, +\\ 
& + & \int_{L_c}^{L_{cut}}W_2(\ell')
n_{pri}(\ell',t)\ell_{pri}(\ell',t)d\ell'\, . \label{ra2tio0}
\end{eqnarray}

The original model for
Goto-Nambu strings included an averaged evolution equation for the
length $\ell$ of each loop, which made the above calculation relatively
easy. Here, an analogous averaged equation for a superconducting loop is
presently unavailable,
but the loop size (and velocity) can be determined by evolving the microscopic
equation of motion (\ref{chflexp}). The functions $W_1(t')$ and $W_2(\ell')$
are `window' functions---typically combinations of Heaviside
functions---selecting the time interval in the evolution of the
network (and the interval in the length of Vachaspati-Vilenkin loops) which
will produce vortons, according to the particular criterion that one chooses to
impose. Notice that these will depend on a number of parameters, including the
initial conditions of the cosmic string network. Also, they should in principle
include a factor accounting for the fact that it takes some time for each
loop to reach a vorton configuration (that is, even if a given loop will
eventually form a vorton, it should not be included in the vorton density
until some time after it is `chopped off' from the long-string network).
However, note that figure \ref{chirsmall} seems to indicate that
this evolution, if
it happens at all, is quite fast---it takes less than a Hubble time.

Note that although in the evolution of the loops the effects of
the currents are
properly accounted for (with the exception of radiative mechanisms), the
evolution of the long string network doesn't take into account of possible
effects of the build-up of the currents. Still, we expect the neglect of these
effects to be a reasonable assumption. This is because such effects should only
become important (if ever) at late times when the network has had time to
build up large currents while, as we will shortly see, most of the energy
density in vortons is produced fairly soon after the network forms (but a
possible exception to this can occur if there are background magnetic fields
which can increase the current build-up rate).

Thus calculation of vorton densities is a two-stage process.
Firstly, one must study the microphysics of the particular model that one is
interested in, in order to derive its microscopic equations of motion and
in particular to construct appropriate expressions for the `window functions'
$W_1(t')$ and $W_2(\ell')$ which will determine at which stages of the
evolution of the string network one can form vortons. Secondly, one
can use the velocity-dependent one-scale model and the model for the
evolution of the currents on the long strings, together with the microscopic
loop equations of motion (or an averaged version of them) to determine the
vorton density using (\ref{ra2tio0}). Typically there will be a single time
interval $t_c\le t_{start}<t<t_{stop}\le t_\star$ at which vortons will form,
but it is relatively easy to think of initial conditions for which vortons
form at two different time intervals. Values of $t_{start}$ and $t_{stop}$
in specific models will be discussed in a forthcoming
publication \cite{inprep}.

Since, as we pointed out, there is some uncertainty in some crucial parameters
of this model, we will limit ourselves in this paper to calculate the vorton
density for GUT and electroweak string networks in the `best' (or `worst'
according to opinion) possible case where there is no equilibration (that is,
$f=0$), the
string-forming and superconducting phase transitions are both of second order,
and all the loops produce vortons (hence our criterion is simply
${\overline n}_{free}>0$). Notice that this last condition is unrealistic for
GUT networks (where, as we already indicated, we don't expect vortons to from)
but is plausible for electroweak networks. Still, we will assume that vortons
can only be formed while the network in in the friction-dominated regime.
Also, since one presumably needs to have quite efficient radiation mechanisms
for all loops to relax into vortons, we will assume that such relaxation is
instantaneous---thus $W_2=1$, while $W_1$ is unity in the friction-dominated
epoch and vanishes afterwards.

\begin{figure}
\vbox{\centerline{
\epsfxsize=0.7\hsize\epsfbox{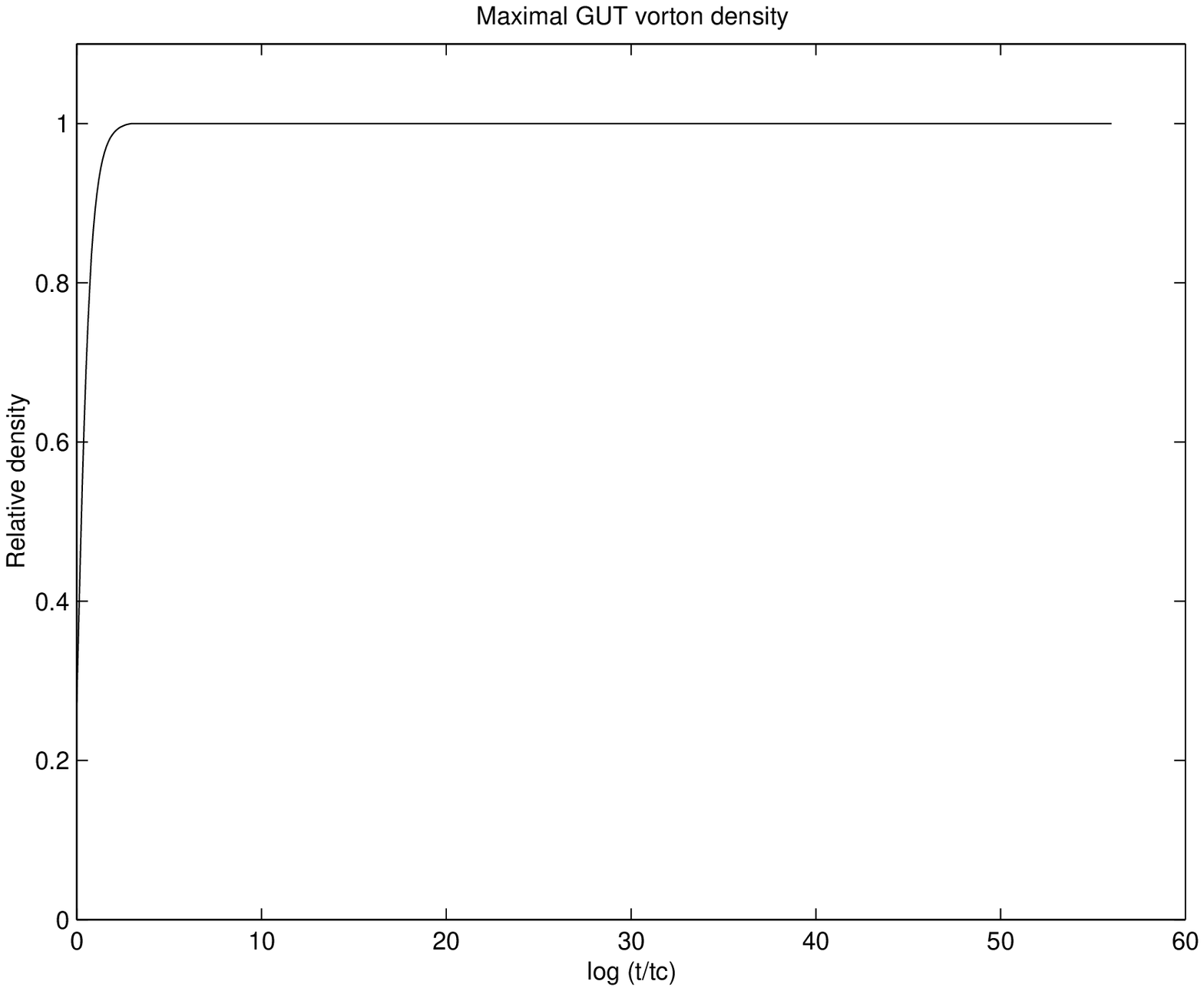}}
\vskip.4in}
\vbox{\centerline{
\epsfxsize=0.7\hsize\epsfbox{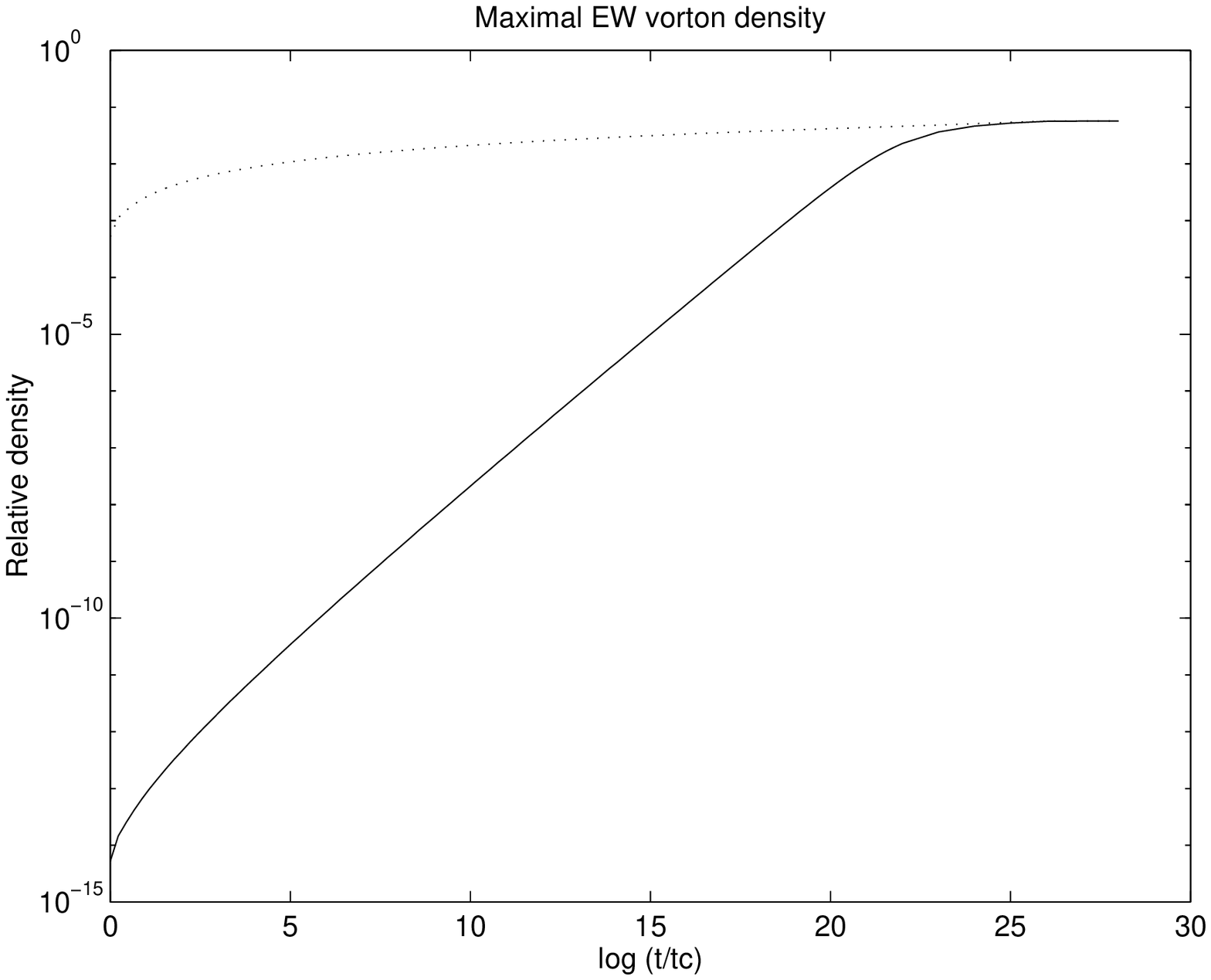}}
\vskip.4in}
\caption{The maximum possible vorton densities relative to the background
(solid lines) and ordinary matter (dotted lines) densities, for GUT and
electroweak-scale string networks. Time is in orders of magnitude from the
epoch of string formation; the plots end at the present epoch.}
\label{vorden}
\end{figure}

Figure \ref{vorden} displays the resulting vorton densities, relative to the
background and matter densities. Firstly, we confirm that most of the
energy density in vortons is produced soon after the network forms. In the
case of GUT strings, we see that vortons would only dominate the energy
density of the universe about four  orders of magnitude in time after the epoch
of network formation, that is soon after friction-domination ends (recall that
for a GUT network $t_\star\sim850t_c$). Thus even if all these vortons formed,
they wouldn't contradict the standard cosmological scenario provided that
they decayed soon after $t_\star$, when the network becomes free. In any
case we emphasise that this `worst case' scenario is not realistic for
GUT-scale strings, and indeed (as discussed previously) we do not expect
GUT-scale vortons to form at all.

On the other hand, electroweak string networks are friction-dominated until
after the radiation-matter transition, so the vorton density has been slowly
building up relative to that of matter until very recently. We find that
this density today would be about $6\%$ of the critical density. On the other
hand, a string network formed at $T\sim10^4\,GeV$ would provide a maximal
vorton density equal to the critical density. This is therefore the strongest
possible vorton constraint---it is based on the assumption that all loops form
vortons.
Naturally, realistic models are not expected to be fully efficient in
producing vortons, and furthermore the relevant phase transitions are not
necessarily of second order. One can therefore conjecture that the dark
matter problem might be solved by a superconducting string network formed
at an energy scale of $T\sim10^5-10^6\,GeV$. Note that there are a number
of super-symmetric models producing such networks (see for
example \cite{riotto}).
We will present a more detailed analysis of these issues in a future
publication \cite{inprep}. 

\section{Conclusions}
\label{v-cc}
In this paper we have presented the first rigourous study of
the cosmological evolution of superconducting
strings in the limit of chiral currents. We have shown that
in this limit the elastic string model of Carter \& Peter\cite{wcp}
coincides with the model derived from first principles by Witten\cite{witten}.

By analysing physically relevant loop solutions of the microscopic
equations of motion for these strings, we have verified that the effect
of frictional damping is crucial for vorton formation. We then defined suitable
parameters characterising the evolution of these loops, and in particular
whether or not they become vortons. In particular, we have established
the usefulness of the `stability parameter' ${\overline n}$.
In general, it is more difficult to form vortons when the
string-forming phase transition is of first order. This is because such
networks produce, during their evolution in the stretching regime, loops with
a size close to that of the horizon; these will therefore be significantly
affected by expansion, which tends to decrease the fraction of the loops's
energy in the current---whereas friction tends to increase it.

After introducing a simple `toy model' for the evolution of currents on the
strings \cite{mss}, we have considered the cases of first and second-order
GUT-scale string-forming and superconducting phase transitions (which is the
most favourable GUT case of vorton formation since frictional forces can act
longer). We have presented evidence suggesting that GUT-scale string networks
might well produce no vortons, and that even if they do, this will not
necessarily rule out such models. This is in contradiction with previous, less
detailed studies \cite{vor,vor2}, and hence calls for a re-examination of a
number of cosmological scenarios involving superconducting strings. Notably, these
strings could be at the origin of
the observed galactic magnetic fields \cite{msmag}.

Finally, we have explicitly calculated the vorton density in two
`extreme' cases
to illustrate the method that one should follow once the microphysical
properties of these networks are known in more detail. For electroweak-scale
string networks, we have found that vortons can produce up to about $6\%$
of the critical density of the universe. On the other hand, it is
conceivable that superconducting string networks formed at an energy
scale $T\sim10^4-10^6\,GeV$ (depending on details of the model) can solve the
dark matter problem.

The detailed analysis presented in this paper for GUT stings can obviously
be extended to other energy scales---this will
be the subject of a forthcoming publication \cite{inprep}. Obviously, as we
lower the energy scale, the frictional force becomes more and more important and
acts for a longer time. Hence the vorton-forming region of parameter space
increases, and by the electroweak scale almost all loops chopped off the
long-string network will become vortons. We therefore conclude that in addition
to the low-$G\mu$ regime (which as we saw includes the electroweak scale)
where vortons can be a source of dark matter and to an
intermediate-$G\mu$ range in which vortons would be too massive to be
compatible with standard cosmology (thereby excluding these models),
there is also a high-$G\mu$ regime (of which the GUT scale is part) in which
vortons don't form at all and therefore no cosmological
constraints based on them can be set.
It is then curious (to say the least) that vorton constraints can be used to
rule out cosmic string models in a wide range of energy scales $G\mu$, but not
those formed around the GUT or the electroweak scales, where cosmic strings
can be cosmologically useful.

\acknowledgments
C.M.\ is funded by JNICT
(Portugal) under `Programa PRAXIS XXI' (grant no.
PRAXIS XXI/BD/3321/94). E.P.S.\ is funded by PPARC and
we both acknowledge the support of PPARC and the
EPSRC, in particular the Cambridge Relativity rolling
grant (GR/H71550) and a Computational Science
Initiative grant (GR/H67652).

\vfill\eject

\vfill\eject

\begin{figure}
\caption{The evolution of chiral circular GUT-scale  string loops formed
at $t=t_c$. All loops have an initial total energy $E_{tot}/2\pi\mu t_c=10$,
but
different initial string energies---respectively $9.3$, $7.2$, $5.0$, $2.8$
and $0.7$; in (a) and (d), these are respectively shown in solid, dashed,
dash-dotted, dotted and starred lines, while in (b) and (c) they are
correspondingly shown by lighter shades of gray. Plots show the total energy
$E_{total}$ relative to $t_c$ (a), the microscopic velocity (b),
the fraction of
the loop's energy in the current (c) and the parameter ${\protect\overline n}$
defined in \protect\ref{ansatznbar} (d).}
\label{chirrad}
\end{figure}

\vfill\eject

\begin{figure}
\caption{The evolution of chiral circular GUT-scale  string loops formed
at $t=t_{\protect\star}$. All loops have an initial total energy
$E_{tot}/2\pi\mu t_{\protect\star}=10$,
but different initial string energies---respectively $9.3$, $7.2$, $5.0$, $2.8$
and $0.7$; in (a) and (d), these are respectively shown in solid, dashed,
dash-dotted, dotted and starred lines, while in (b) and (c) they are
correspondingly shown by lighter shades of gray. Plots
show the total energy $e_{total}$ relative to $t_{\protect\star}$ (a), the microscopic
velocity (b), the fraction of the loop's energy in the current (c) and the
parameter ${\protect\overline n}$ defined in \protect\ref{ansatznbar} (d).}
\label{chirfri}
\end{figure}

\vfill\eject

\begin{figure}
\caption{The evolution of chiral circular GUT-scale string loops formed
at $t=t_c$. All loops have an initial string energy $E_{string}/2\pi\mu t_c=0.1$,
but different initial ratios of energies in the current and the
string---the cases $10^{-3}$, $10^{-2}$, $10^{-1}$, $1.0$ and $2.0$ are
respectively shown in solid, dashed, dash-dotted, dotted and starred
lines. Plots show the total energy $E_{total}$ relative to $t_c$ (a), the (base-ten)
logarithm of the microscopic velocity (b), the fraction of
the loop's energy in the current (c) and the (base-ten) logarithm of the
parameter $1-{\protect\overline n}$ (d).}
\label{chirsmall}
\end{figure}

\vfill\eject

\begin{figure}
\caption[Maximum velocity of chiral circular loops]{The maximum microscopic
velocity reached by circular GUT-scale
chiral superconducting string loops formed at a time
$t_f=t_c\protect\sim10^{-39} s$ (top left) and $t_f=10, 100, 855\,t_c$ (clockwise),
the later being $t_\protect\star$ The `x' axis corresponds to the initial
value of the parameter ${\overline n}$, going from zero (the Goto-Nambu case)
to unity; in the `y' axis the base-ten log of the string radius relative 
to the horizon size goes from $-2$ to $1$. Note that in the first graph the
friction lengthscale corresponds to
$\protect\log{\protect\ell_{\protect\rm f}/t}\protect\sim-1.5$, while in the
last one
$\protect\log{\protect\ell_{\protect\rm f}/t}\protect=0$.}
\label{chmaxv}
\end{figure}

\vfill\eject

\vfill\eject

\vfill\eject

\vfill\eject

\vfill\eject

\vfill\eject

\vfill\eject

\end{document}